\documentclass[traditabstract]{aa}
\usepackage{graphicx}
\usepackage{txfonts}
\usepackage{natbib}
\usepackage{multirow}
\usepackage{longtable}
\usepackage{lscape}
\usepackage{url}
\bibpunct{(}{)}{;}{a}{}{,} 

\begin{document}
	\title{Dynamical analysis of strong-lensing galaxy groups at intermediate redshift\thanks{Based on observations collected at the European Southern Observatory, Paranal, Chile (Program P80.A-0610B); based on observations obtained with MegaPrime/MegaCam, a joint project of CFHT and CEA/DAPNIA, at the Canada-France-Hawaii Telescope (CFHT) which is operated by the National Research Council (NRC) of Canada, the Institut National des Sciences de l'Universe of the Centre National de la Recherche Scientifique (CNRS) of France, and the University of Hawaii. This work is based in part on data products produced at TERAPIX and the Canadian Astronomy Data Centre as part of the Canada-France-Hawaii Telescope Legacy Survey, a collaborative project of NRC and CNRS.}}
	\titlerunning{Dynamical analysis of SL2S galaxy groups}
	\authorrunning{R.P. Mu\~noz et al.}

   \author{R.P. Mu\~noz \inst{1,2} \and V. Motta \inst{1} \and T. Verdugo \inst{1,3} \and F. Garrido \inst{2} \and M. Limousin \inst{4,5} \and N. Padilla \inst{2}  \and G. Fo\"ex \inst{1} \and R. Cabanac \inst{6} \and \\ R. Gavazzi \inst{7} \and L.F. Barrientos \inst{2} \and J. Richard \inst{8}  }

   \institute{Departamento de F\'isica y Astronom\'ia, Universidad de Valpara\'iso, Avda. Gran Breta\~na 1111, Valpara\'iso, Chile e-mail: rmunoz@astro.puc.cl \and Departamento de Astronom\'ia y Astrof\'isica, Pontificia Universidad Cat\'olica de Chile, Avda, Vicu\~na Mackenna 4860, Casilla 306, Santiago 22, Chile \and Centro de Investigaciones de Astronom\'ia, AP 264, M\'erida 5101-A, Venezuela \and Aix Marseille Universit\'e, CNRS, LAM (Laboratoire d'Astrophysique de Marseille) UMR 7326, 13388, Marseille, France
    \and Dark Cosmology Centre, Niels Bohr Institute, University of Copenhagen, Juliane Marie Vej 30, 2100 Copenhagen, Denmark \and Laboratoire d'Astrophysique de Toulouse-Tarbes, Universit\'e de Toulouse, CNRS, 57 avenue d'Azereix, 65000 Tarbes, France \and Institut d'Astrophysique de Paris, UMR 7095  CNRS \& Universit\'e Pierre et Marie Curie, 98bis Bd Arago, F-75014 Paris, France \and CRAL,  Universit\'e Lyon 1, Observatoire de Lyon, 9 avenue Charles Andr\'e, 69561 Saint Genis Laval Cedex, France}

   \date{Received ; accepted}


  \abstract
   {We present VLT spectroscopic observations of 7 discovered galaxy groups between $0.3<z<0.7$. The groups were selected from the Strong Lensing Legacy Survey (SL2S), a survey that consists in a systematic search for strong lensing systems in the Canada-France-Hawaii Telescope Legacy Survey (CFHTLS). We give details about the target selection, spectroscopic observations and data reduction for the first release of confirmed SL2S groups. The dynamical analysis of the systems reveals that they are gravitationally bound structures, with at least 4 confirmed members and velocity dispersions between 300 and $800\;\mathrm{km}\,\mathrm{s}^{-1}$. Their virial masses are between $10^{13}$ and $10^{14}\;M_\odot$, and so  can be classified as groups or low mass clusters. Most of the systems are isolated groups, except two of them that show evidence of an ongoing merger of two sub-structures. We find a good agreement between the velocity dispersions estimated from the analysis of the kinematics of group galaxies and the weak lensing measurements, and conclude that the dynamics of baryonic matter is a good tracer of the total mass content in galaxy groups.}


   \keywords{Galaxies: groups: general --
     Galaxies: kinematics and dynamics --
     Galaxies: distances and redshifts
         }

   \maketitle
%

\section{Introduction}

Galaxy groups are the most common structures in the Universe, containing at least 50\% of all galaxies at the present day \citep{eke04a}, and cover the intermediate mass range between large elliptical galaxies and galaxy clusters. 
 A wide array of methods have been used to identify groups at intermediate and high-z: Percolation algorithms based on optical photometric and spectroscopic data \citep{mari02,eke04a,adam05,yan07,zap09}, X-ray emission from hot intragroup gas \citep{bro06,fino09}, bright arcs due to strong lensing \citep{cab07,lim09,mor12}.


The first large sample of groups detected in redshift space was presented by \citet{gel83}, who found 176 groups up to $z=0.03$ by using the Center for Astrophysics (CfA) galaxy redshift survey. Nowadays, with the advent of large spectroscopic surveys such as the Two Degree Field Galaxy Redshift Survey \citep[2dFGRS;][]{col01}, the Sloan Digital Sky Survey \citep[SDSS;][]{yor00}, and the Deep Extragalactic Probe 2 Redshift Survey \citep[DEEP2;][]{dav03}, well over 5,000 groups have been identified up to $z\sim1$. \citet{eke04a} identified about $7,000$ groups and clusters in the 2dFGRS with at least 4 members, and found that they cover a wide range in mass between $10^{12}$ and $10^{15}\;M_\odot$. At low redshift, \citet{berl06} used a friends-of-friends (FOF) algorithm to identify groups in the SDSS Data Release 2 \citep[DR2;][]{aba04} and found about $8,100$ groups between $0.01<z<0.10$. At higher redshift, \citet{ger05} found about 900 groups with 2 or more members between $0.7<z<1.4$ using the DEEP2 survey.

Recently, \citet{knob12} presented a sample of about $1,500$ galaxy groups between redshifts 0.1 and 1.0 that were identified in the zCOSMOS-bright survey \citep{lilly07}. They detected a clear evidence for the growth of cosmic structure over the last seven billion years because the fraction of galaxies that are found in groups (in volume-limited samples) decreases significantly to higher redshifts.

The analysis of the galaxy content in groups and clusters is essential to understand the effects of the local environment on galaxy formation and evolution processes. For instance, galaxy collisions are expected to be most effective in less massive groups, where the system velocity dispersions are comparable to the internal velocities of galaxies, leading to strong galaxy-galaxy interactions and therefore enhancing star formation \citep{mih96}. Furthermore, the feedback from supernovae and supermassive black holes on the hot intragroup gas is expected to be relevant in suppressing the onset of catastrophic cooling of the hot gas \citep[eg.][]{chu01}, since the energy input associated with these sources is comparable to the binding energies of these systems \citep{mcc10}.

The evolution of the stellar and gas content of galaxies strongly depends on the properties of their host galaxy cluster \citep{oem74}. For instance, \citet{han09} studied a large sample of groups and clusters in the Sloan Digital Sky Survey \citep[SDSS;][]{yor00}, and found that the fraction of red-sequence galaxies increases with cluster mass and it decreases with cluster-centric distance (see also \citet{pad10}).

{
Ir order to obtain reliable conclusions about the relative importance of different physical processes in driving galaxy evolution in groups, it is necessary to build composite samples of galaxy groups and then use the virial mass as the mass normalization.
\citet{biv06} studied the accuracy of the virial mass estimate from numerical simulations
using both the dark-matter particles and simulated galaxies in $67$ synthetic clusters.
To analyze how the observational strategy and sample sizes affect the cluster mass estimates, they used these synthetic clusters to select a sample of galaxies and estimate the cluster dynamical mass by using two different estimators..
They found that the total mass of clusters can be estimated with an accuracy of $10$ to $15$ percent
when using $400$ cluster members, and that these figures become twice as large when
the available number of members is $20$. In this paper, we follow \citet{biv06} to analyze our lens galaxy groups.
}

In this paper we introduce the first spectroscopically confirmed groups of the SL2S survey that were observed at the Very Large Telescope (VLT). {The SL2S survey is a small and well defined sample of groups of galaxies selected by their strong lensing features \citep{cab07,mor12}.}
 The description of the  spectroscopic observations, their reduction and calibrations are presented in Section \ref{sec:obs}. In Section \ref{sec:analysis} we present the membership determination and velocity dispersion estimation for the confirmed groups. The numerical simulations used to study the accuracy of the velocity dispersion estimations are presented in Section \ref{sec:sim}. The discussion of the main results is presented in Section \ref{sec:discussion}, and the conclusions are summarized in Section \ref{sec:conclusions}.

We assume a $\Lambda$CDM cosmology with $\Omega_M=0.25$, $\Omega_{\Lambda}=0.75$, and a Hubble constant of $H_0=73\;\mathrm{km}\,\rm{s}^{-1}\,\rm{Mpc}^{-1}$.

\section{Observations and Data Reduction}
\label{sec:obs}

The groups studied in this work were selected from the Strong Lensing Legacy Survey \citep[SL2S;][]{cab07}, a large systematic search for strong-lensing systems in the Canada-France-Hawaii Telescope Legacy Survey (CFHTLS)\footnote{\url{http://www.cfht.hawaii.edu/Science/CFHTLS}}. The detection and classification of group candidates is explained in \citet{cab07}, but it basically consisted in running the ARCFINDER algorithm by \citet{ala06} on the stacked CFHTLS images and then doing a visual inspection to reject spurious candidates.

Recently, \citet{mor12} has published a catalog of 127 strong-lensing systems detected in the SL2S survey with photometric redshifts between $0.2$ and $1.2$.  They found a systematic alignment of the giant arcs with the major axis of the baryonic component of the putative lens, and more important, they were able to probe the average density profiles of groups using the image separation distribution. Several SL2S systems presented in \citet{cab07} and \citet{mor12} have been followed-up with optical observations at the Hubble Space Telescope (HST), near-infrared observations at the CFHT and optical spectroscopy at the ESO Very Large Telescope (VLT).

In this work, we present medium-resolution spectroscopy of 8 SL2S systems observed at the ESO VLT telescope. They were selected for showing extended arcs with Einstein radius ($R_E$) lower than $8\;\arcsec$, and having photometric redshifts ($z_{phot}$) between 0.3 and 0.7. The selection criteria is based on the predicted angular separations from N-body numerical simulations of dark matter haloes by \citet{ogu06}, where they obtained that strong-lensing arcs with $3\arcsec < R_E < 8\arcsec$ are likely generated by galaxy-group scale dark matter haloes.

Several of the systems presented in this work have weak lensing mass estimates from \citet{lim09}. They measured the weak lensing signal for 13 SL2S systems between $0.3<z_{phot}<0.8$, and were able to estimate weak lensing masses for 6 of them. Furthermore, the gravitational potential of the system \mbox{SL2S02140-0535} presented in this work was studied in detail by \citet[see][]{ver11} by combining strong-lensing, weak-lensing and dynamic measurements.

\subsection{Imaging}

The groups have $u$, $g$, $r$, $i$ and $z$-band photometry as part of the CFHTLS survey, a major photometric survey in five bands that covers a total area of 159 $\rm{deg}^2$ (T0006 release of the CFHTLS Deep and Wide surveys; see more details in Goranova et al. 2009 \footnote{\url{http://terapix.iap.fr/cplt/T0006-doc.pdf}}). The CFHTLS survey observations were obtained at the 3.6m CFHT telescope with the MEGACAM camera, a wide field imager that consists of 36 2048x4612 pixels CCDs of pixel size $0.186\arcsec$.


The images and photometric catalogs used in this work are based on the T0005 release of the CFHTLS survey (November, 2008), and were built at the TERAPIX data processing center at the Institut d'Astrophysique de Paris (IAP; see more details in Mellier et al. 2008 \footnote{\url{http://terapix.iap.fr/cplt/oldSite/Descart/CFHTLS-T0005-Release.pdf}}). The 50\% completeness magnitude of point-like sources in these catalogs are $u=25.34$, $g=25.47$, $r=24.82$, $i=24.48$ and $z=23.60$.



\subsection{Spectroscopy}


We obtained medium resolution spectra of group galaxies with the Focal Reducer and low dispersion Spectrograph 2 \citep[FORS2;][]{app98} at the VLT telescope. The FORS2/VLT observations were carried out during the ESO observing programme P80.A-0610B (P.I. Motta) and consisted of multi-object spectroscopy (MOS) of 8 SL2S systems. We used a medium resolution grism (GRIS\_600RI, 0.83 \AA/pix) since we wanted to measure the internal velocity dispersion of the brightest group galaxies, and adopted a $2\times2$ binning in order to improve the signal-to-noise ratio (SNR) of the spectra.

Depending on the number of group member candidates, we used one or two FORS2 masks to do MOS of each group. One FORS2 mask allowed us to take spectra of $\sim 40$ targets simultaneously within a field of view of $4.25\arcmin\times4.25 \arcmin$. The criteria used to select the galaxies that entered in the MOS masks was based on the magnitudes and colors of galaxies. We defined as candidates those galaxies with magnitudes $i<22.0$ and colors within $(g-i)_{lens}-0.15 < g-i < (g-i)_{lens}+0.15$, where $(g-i)_{lens}$ is the color of the brightest lens galaxy within the $R_E$. As the masks could not be filled only with group candidates, we randomly selected galaxies within the field of view with $i<20.0$ and no color restrictions.

For all the masks we obtained two exposures of 1400 s each, except for the SL2SJ08591-0345 mask where we used only one exposure because of time constraints. We found that two exposures were enough to remove most of the cosmic rays from the 2D spectra, although a couple of them were not removed and had to be manually masked in the 1D spectra. The number of masks and total exposure time for each group is given in Table \ref{tab:observations}.

The MOS masks were reduced using the standard ESO data reduction procedures\footnote{Very Large Telescope Paranal Science Operations FORS data reduction cookbook, v1} and the Optimal Spectrum Extraction Package (OSEP) for IDL\footnote{\url{http://physics.ucf.edu/~jh/ast/software/optspecextr-0.3.1/doc}}. The basic data reduction steps consisted on bias subtraction, flat-fielding and wavelength calibration, which were done using the ESO Recipe Execution Tool (EsoRex; \url{http://www.eso.org/sci/software/cpl/esorex.html}) and the Common Pipeline Library (CPL; \url{http://www.eso.org/sci/software/cpl}). The advanced steps consisted on the removal of cosmic rays, the background subtraction from the 2D spectra, the 1D spectra extraction and the average of multiple spectra for each source, and they were done using the OSEP IDL procedures inspired in the optimal extraction algorithm by \citet{hor86}.

\begin{table}
\caption{VLT/FORS2 spectroscopic data}
\label{tab:observations}
\centering
\begin{tabular}{l @{}c@{} c c c}
\hline\hline
Target & \phantom{xxx} & Date & Mask & Exposure  \\
\hline
SL2SJ02132-0743 & & 2008-08-11 & M013 & $2\times1400$  \\
\multirow{2}{*}{SL2SJ02140-0535} & \multirow{2}{*}{\bigg\{} & 2007-10-19 & M012 & $2\times1400$  \\
                   & & 2008-02-01 & M010 & $2\times1400$  \\
SL2SJ02141-0405 & & 2007-10-19 & M014\_1 & $2\times1400$  \\
\multirow{2}{*}{SL2SJ02180-0515} & \multirow{2}{*}{\bigg\{} & 2008-02-07 & M014\_2 & $2\times1400$  \\
                   & & 2008-08-26 & M014\_3 & $2\times1400$  \\
\multirow{2}{*}{SL2SJ02215-0647} & \multirow{2}{*}{\bigg\{} & 2008-08-24 & M016\_1 & $2\times1400$  \\
                   & & 2008-09-10 & M016\_2 & $2\times1400$  \\
\multirow{2}{*}{SL2SJ08544-0121} & \multirow{2}{*}{\bigg\{} & 2007-12-12 & M005 & $1\times1200$  \\
                   & & 2008-02-06 & M005 & $2\times1400$  \\
\multirow{2}{*}{SL2SJ08591-0345} & \multirow{2}{*}{\bigg\{} & 2007-12-12 & M002\_1 & $1\times1400$ \\
                   & & 2007-12-18 & M002\_2 & $1\times1400$ \\
SL2SJ09413-1100 & & 2007-12-18 & M001 & $2\times1400$  \\
\hline
\end{tabular}
\tablefoot{
The columns show the name of the target; UT date of observations; name of the multi-object spectroscopy mask; number of individual exposures and their corresponding exposure time in seconds.
}
\end{table}


\section{Analysis \& Results}
\label{sec:analysis}

\subsection{Redshift measurements}
\label{sec:redshift}

The spectroscopic redshifts were determined using the Radial Velocity SAO package \citep[RVSAO;][]{kur98} within the IRAF software\footnote{IRAF is distributed by the National Optical Astronomy Observatories, which are operated by the Association of Universities for Research in Astronomy, Inc., under cooperative agreement with the National Science Foundation.}. We first identified several emission and absorption lines by doing visual inspection of the galaxy spectra, and then we determined the redshifts by cross-correlating a spectrum against template spectra of known velocities.  

The galaxy spectra cover the wavelength range $5200\;\AA < \lambda < 8400\;\AA$, and the SNR per resolution element varies from $\sim5$ to $\sim30$. For most of the galaxy spectra we were able to identify the \textsc{Ca II K+H},  \textsc{G}-band and \textsc{Mg I} absorption lines, and for few of them we also identified the \textsc{O II}, $\textsc{H}\beta$, \textsc{O III} and $\textsc{H}\alpha$ emission lines. The errors in the redshift measurements is affected by the instrumental resolution and the RVSAO template fit, and it is $\delta z =0.001$.

The redshifts were classified in three types: secure, questionable and unknown. Secure redshifts correspond to spectra having at least three identified lines, between absorption and emission lines; questionable redshifts to spectra having only one or two identified lines; and unknown redshifts to spectra having no identified lines. The success ratio of secure redshift determination is between 50\% and 70\%, and correspond to groups SL2SJ02140-0535 and SL2SJ08544-0121, respectively. We found that this ratio strongly depends on the total exposure time and magnitude of the targets.

For all the SL2S group candidates we were able to measure the redshift of the brightest galaxy within the Einstein radius, hereafter called main lens galaxy. The spectra of the main lens galaxy of SL2S groups are shown in the top panel of figures in Appendix \ref{sec:ap1}, and the main absorption and emission lines have been identified.

\begin{figure*}[!htp]
\centering
\includegraphics[width=\textwidth]{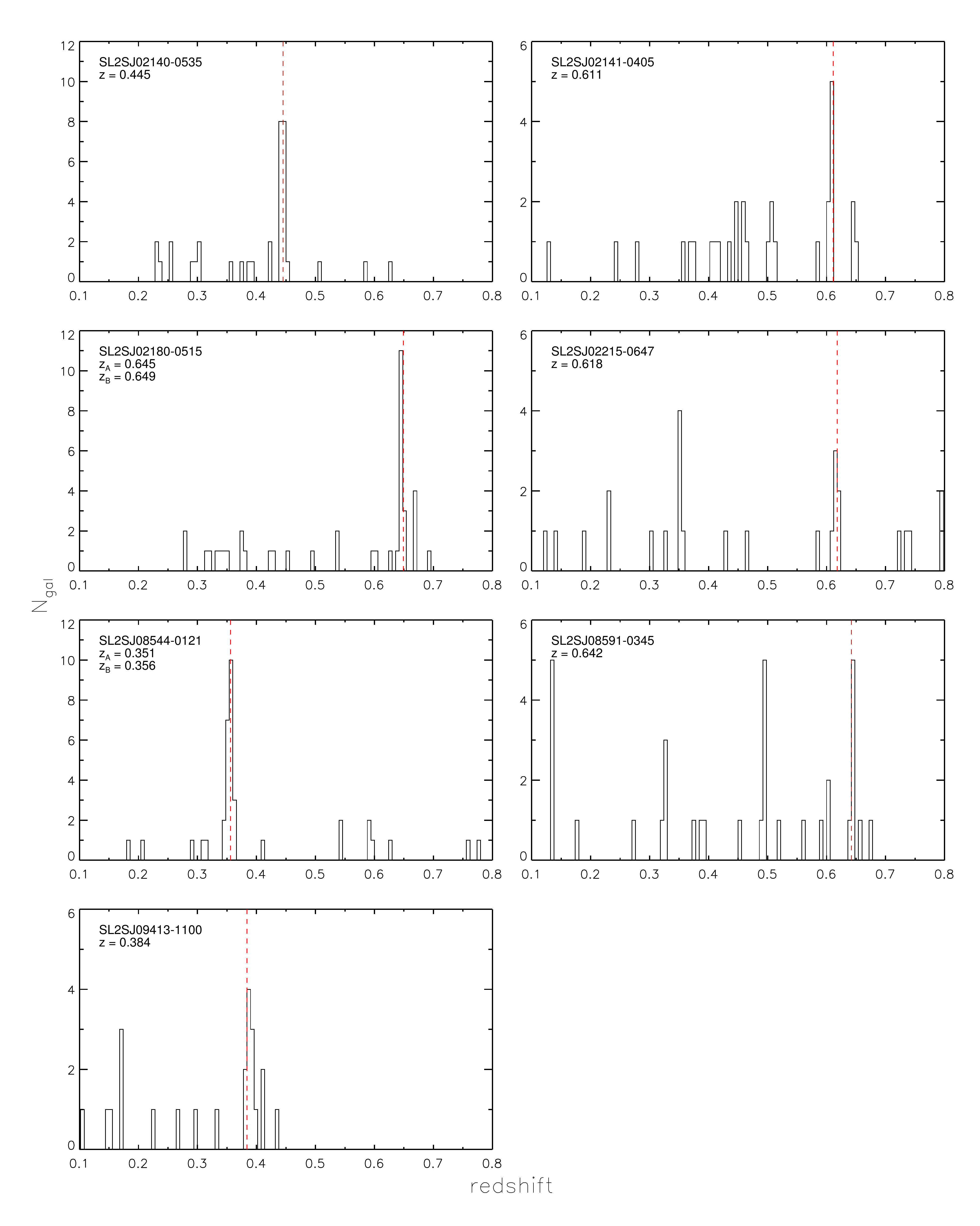}
\caption{Redshift distribution of galaxies in the group fields. The spectroscopic redshift of isolated groups is denoted by the quantity $z$, and for double components of bimodal groups, by $z_A$ and $z_B$. The vertical dashed line corresponds to the redshift of the most massive component. The bin size of the histogram is $\delta z=0.006$ }
\label{fig:distribution}
\end{figure*}

The galaxy redshift distributions in the direction of the SL2S group candidates are shown in Figure \ref{fig:distribution}. For five of the eight group candidates, we detected a strong peak in the redshift distribution around the redshift of the main lens galaxy. The red dashed lines correspond to the redshift of the group center of mass (see Section \ref{sec:velocity}).






\subsection{Group membership and velocity dispersions}
\label{sec:velocity}

We adopted the formalism by \citet{wil05} to determine the group membership of the SL2S systems. For the systems that showed a single peak in the redshift distribution around the redshift of the main lens galaxy, we identified the group members as follows: the group was initially assumed to be located at the redshift of the main lens galaxy, $z_{lens}$, with an initial observed-frame velocity dispersion of $\sigma(v)_{obs} = 500(1+z_{lens})\;km\,s^{-1}$. Then, we computed the maximum redshift shell, $\delta z_{max}$, and the maximum spatial distance, $\delta \theta_{max}$, as following,

\begin{eqnarray}
\delta z_{max} & = & \frac{2 \sigma(v)_{obs}}{c} \;,\\
\delta \theta_{max} & = & 206,265'' \frac{c\, \delta z_{max}}{b(1+z_{lens})H(z)D_\theta (z)} \;,
\end{eqnarray}

\noindent where c is the speed of light, $H(z)$ is the Hubble constant at z, $D_\theta(z)$ is the angular diameter distance at z, and b is the axis ratio of the cylindrical linking volume. { In N-body numerical simulations of dark matter halos, the cylindrical linking volume is a cylinder oriented along the line of sight with a radius equal to the projected linking length.} We adopted a value of b=3.5 in this work.

The initial guess for the velocity dispersion was based on the estimated velocity dispersion of the strong-lening galaxy group B2108+213 measured by \citet{mck10}. They obtained a mean value of $555\;km\,s^{-1}$ for the galaxy group using three different linking velocity kernels. The factor $(1+z_{lens})$ is used to account for the cosmlogical expansion of the Universe.

Upon identifying potential members as those galaxies located inside the maximum redshift shell and the maximum projected distance, we computed the observed velocity dispersion of the group $\sigma(v)_{obs}$. For those groups with more than 10 members we used the biweight estimator \citep{bee90} in order to compute the velocity dispersion, and for those with less than 10 members we used the gapper algorithm \citep{bee90}. The new computed value of $\sigma(v)_{obs}$ was then used to compute new values of $\delta z_{max}$ and $\delta \theta_{max}$. Finally, we defined as confirmed group members those galaxies located within these limits.

\begin{table*}
\caption{Summary of confirmed SL2S groups.} \label{tab:summary}
\begin{center}
\begin{tabular}{l @{}c@{} l l l l r r r }
\hline\hline
\multicolumn{1}{c}{Group} & \phantom{xxx} & \multicolumn{1}{c}{RA} & \multicolumn{1}{c}{DEC} & \multicolumn{1}{c}{$\bar{z}_{spec}$} & \multicolumn{1}{c}{$\sigma (\rm{v})_{los}$} & \multicolumn{1}{c}{$N_{spec}$} & \multicolumn{1}{c}{$N_{spec,RS}$} & \multicolumn{1}{c}{$N_{gal}$} \\
\hline
SL2SJ02140-0535 & & 02:14:08.03 & -05:35:32.3 & 0.445 & $364^{+60}_{-137}$ & 16 & 11 & 40 \\
SL2SJ02141-0405 & & 02:14:11.21 & -04:05:02.8 & 0.611 & $478^{+48}_{-178}$ & 7 & 5 & 6 \\
\multirow{2}{*}{SL2SJ02180-0515} & \multirow{2}{*}{\bigg\{} & 02:18:10.09 & -05:15:33.5 & 0.645 & $131^{+29}_{-61}$ & 6 & \multirow{2}{*}{8} & \multirow{2}{*}{8} \\
 & & 02:18:07.29 & -05:15:36.2 & 0.649 & $148^{+37}_{-56}$ & 5 & & \\
SL2SJ02215-0647 & & 02:21:51.17 & -06:47:33.7 & 0.618 & $234^{+76}_{-40}$ & 4 & 2 & 2 \\  
\multirow{2}{*}{SL2SJ08544-0121} & \multirow{2}{*}{\bigg\{} & 08:54:46.55 & -01:21:36.9 & 0.351 & $185^{+30}_{-62}$ & 8 & \multirow{2}{*}{10} & \multirow{2}{*}{64} \\   
 & & 08:54:47.10 & -01:21:35.6 & 0.356 & $341^{+43}_{-109}$ & 10 &  &  \\
SL2SJ08591-0345 & & 08:59:14.46 & -03:45:14.2 & 0.642 & $507^{+107}_{-336}$ & 5 & 3 & 6 \\  
SL2SJ09413-1100 & & 09:41:34.99 & -11:00:55.0 & 0.384 & $350^{+57}_{-210}$ & 5 & - & - \\ 
\hline
\end{tabular}
\end{center}
\tablefoot{The columns show the name of the identified SL2S groups; J2000.0 coordinates of the brightest group member with spectroscopic redshift; spectroscopic redshift of the group, $\bar{z}_{spec}$;  line-of-sight velocity dispersion of the group in units of $km\;s^{-1}$; number of spectroscopically confirmed members, $N_{spec}$; number of confirmed members within the E/S0 ridgeline (RS) and inside a group-centric distance of $1\;h^{-1} Mpc$, $N_{spec,RS}$; estimated total number of group members within the E/S0 ridgeline and inside $1\;h^{-1} Mpc$, $N_{gal}$ (see Section \ref{sec:discussion}).\\}
\end{table*}

For the groups that showed a bimodal redshift distribution, we identified the group members of each component as follows: the first component was initially assumed to be located at the higher redshift peak, $z_{peak,high}$, with an initial observed-frame velocity dispersion of $\sigma(v)_{obs} = 250(1+z_{peak,high})\;km\,s^{-1}$. Then, we applied the same procedure we detailed before for groups with a single component, and finally determined the group members of the higher redshift component.

To determine the membership of the lower redshift component, first we excluded the galaxies linked to the first component. Then, we repeated the same procedure used for the first component, but this time using the lower redshift peak as a first guess.

We classified as groups those structures having at least 4 confirmed members in order to reduce the contamination by spurious structures in the catalogs. The final list of SL2S groups is shown in Table \ref{tab:summary}, and consists of five structures with a single component in redshift space, and two structures with a double component. The system \mbox{SL2SJ02132-0743} was excluded from the group list since only two galaxies have a redshift consistent with the redshift of the main lens galaxy and also no overdensity in redshift space was found.

The group redshifts, $\overline{z}_{spec}$, and the number of spectroscopically confirmed members, $N_{spec}$, are shown in columns 4 and 6 of Table \ref{tab:summary}. The group redshift was computed by taking the mean of the redshift of the spectroscopically confirmed members (presented in Table \ref{tab:mask} of Appendix B), and then removing the peculiar motion of the Sun with respect to the CMB. In order to study the accuracy of the velocity dispersions and mass estimates presented in this work (see Section \ref{sec:mass}), we also computed the number of confirmed members with colors consistent with the observed E/S0 ridgeline of galaxy groups and clusters \citep{dre84} and located inside a group-centric distance of $1\;h^{-1}Mpc$, and denoted it by $N_{spec,RS}$ in Table \ref{tab:summary}.

The group velocity dispersions were estimated using the recessional velocity of their members.
The measured redshift of a galaxy member, $z$, can be related to its peculiar velocity with respect to the group center of mass by the following,

\begin{displaymath}
1+z = (1+z_O)(1+z_R)(1+z_G) \;;
\end{displaymath}

\noindent where $z_O$ is the local observer \textit{O} comoving with the expanding Universe, $z_R$ is the cosmological redshift of the structure as measured by \textit{O} relative to a comoving observer \textit{R} in the vicinity of the structure, and $z_G$ is the peculiar velocity of the galaxy with respect to the center of mass of the structure. The $z_O$ contribution is negligible for groups at $z>0.02$.

Substituting $z_R$ by the spectroscopic redshift of the group, we obtain the following equation for the line-of-sight velocity of a galaxy with respect to the group center of mass,

\begin{eqnarray}
v_{los} &=& \frac{c(z-\bar{z}_{spec})}{1+\bar{z}_{spec}} \label{eqn1} \;
\end{eqnarray}

\noindent where $\bar{z}_{spec}$ is the group redshift as shown in Table \ref{tab:summary}.

{
The group velocity dispersion is related to the sum in quadrature of the $v_{los}$ of all group members, and this value is affected by the recessional velocity errors. We computed the rest-frame velocity dispersion of each group by applying the biweight estimator of scale \citep{bee90} to the $v_{los}$ of its members. In order to remove the recessional velocity errors from the velocity dispersion measurements, we followed the prescription by \citet{dan80}, i.e., we subtracted in quadrature the mean $v_{los}$ errors from the velocity dispersion. The corrected line-of-sight velocity dispersions, $\sigma(v)_{los}$, of the SL2S groups are shown in Table \ref{tab:summary}. The upper and lower errors in $\sigma(v)_{los}$ were estimated by using a bootstrap technique of $10,000$ repetitions (see \citet{bee90} for details on the methodology).
}

\subsection{Mass Estimates}
\label{sec:mass}

The total mass of the SL2S groups was estimated by using the virial theorem. We assumed that the groups are in hydrostatic equilibrium, have spherical symmetry and have isotropic velocity distributions. It is important to note that the first assumption could be wrong for the youngest and less massive groups, at which the internal velocity dispersions of the galaxies are comparable to that of the group and, therefore, galaxy mergers are favored \citep{hic97}.

For distant galaxy clusters and groups, it is only possible to measure their projected velocity dispersions and galaxy separations. We computed the projected virial radius for each group using the sky angular distances between all its members, following the formalism by \citet{gir98}. The projected virial radius and mass were computed as follows,

\begin{eqnarray}
R_{PV} & = & D_\theta(\bar{z})  N(N-1) \frac{1}{ \sum_{i=1}^{N} \sum_{j=i+1}^{N} \frac{1}{\theta_{ij}} } \;,\\
M_V & = & \frac{3\pi}{2} \frac{\sigma(v)_{los}^2 R_{PV}}{G} \,,
\end{eqnarray}

\noindent where $R_{PV}$ is the projected virial radius, $M_V$ is the virial mass, $D_\theta(\bar{z})$ is the angular diameter distance at redshift z, $N$ is the number of confirmed members, $\theta_{ij}$ is the sky angular distance between galaxies i and j, and G is the gravitational constant.

The projected virial radii and virial masses of SL2S groups are shown in Table \ref{tab:mass}, in units of Mpc and $10^{14}\;M_\odot$, respectively.

\begin{table}
\caption{Virial radii and masses of SL2S groups.}
\label{tab:mass}
\centering
\begin{tabular}{l @{}c@{} l l l c}
\hline\hline
\multicolumn{1}{c}{Group} & \phantom{xxx} & \multicolumn{1}{c}{$\bar{z}_{spec}$} & \multicolumn{1}{c}{$R_{PV}$} & \multicolumn{1}{c}{$M_{V}$} & \multicolumn{1}{c}{$M_{WL}(2 \rm{Mpc})$ \tablefootmark{a}} \\
\hline
SL2SJ02140-0535 & & 0.445 & 0.78 & $1.14^{+0.41}_{-0.69}$ & $5.5\pm3.7$ \\
SL2SJ02141-0405 & & 0.611 & 0.43 & $1.08^{+0.23}_{-0.66}$ &  \\
\multirow{2}{*}{SL2SJ02180-0515} & \multirow{2}{*}{\bigg\{} & 0.645 & 0.69 & $0.13^{+0.02}_{-0.09}$ & $-$ \tablefootmark{b} \\
 & & 0.649 & 0.23 & $0.05^{+0.01}_{-0.03}$ & $-$ \tablefootmark{b} \\
SL2SJ02215-0647 & & 0.618 & 0.72 & $0.43^{+0.33}_{-0.04}$ & $<3.1$ \tablefootmark{c} \\
\multirow{2}{*}{SL2SJ08544-0121} & \multirow{2}{*}{\bigg\{} & 0.351 & 0.45 & $0.17^{+0.03}_{-0.09}$ & \multirow{2}{*}{$6.3\pm2.5$} \\
 & & 0.356 & 1.07 & $1.37^{+0.37}_{-0.73}$ \\
SL2SJ08591-0345 & & 0.642 & 0.53 & $1.51^{+0.71}_{-1.34}$ & $-$ \tablefootmark{d} \\
SL2SJ09413-1100 & & 0.384 & 1.11 & $1.49^{+0.52}_{-1.26}$ & $3.7\pm3.4$ \\
\hline
\end{tabular}
\tablefoot{
The columns show the name of the SL2S group; spectroscopic redshift of the group, $\bar{z}_{spec}$; projected virial radius of the group, $R_{PV}$, in units of Mpc; virial mass of the group in units of $10^{14}\;M_\odot$; projected mass derived from weak lensing as computed within a circular aperture of radius 2 Mpc, in units of $10^{14}\;M_\odot$.\\
\tablefoottext{a}{$M_{WL}$ inside a projected radius of 2 Mpc taken from \citet{lim09}}\\
\tablefoottext{b}{$R_{E}$ is below 3 arcsec and is within galaxy lensing regime.}\\
\tablefoottext{c}{The weak lensing signal is low, and only an upper limit could be established.}\\
\tablefoottext{d}{The system is located at the edge of the field of view.}
}
\end{table}




\section{Accuracy of velocity dispersion and mass estimates}
\label{sec:sim}

The dynamical masses estimated for the lensed clusters can be subject to statistical
and systematic biases.  In this section we use {\small GALFORM} semi-analytic galaxies from the \citet{bow06} version of the
model, which populate the Millennium simulation \citep{spr05}.

This simulation
adopts a flat $\Lambda$CDM cosmology with $z=0$ dark-matter and baryon density
parameters $\Omega_{dm}=0.205$, $\Omega_b=0.045$, a dimensionless Hubble constant of $h=0.73$,
$rms$ linear mass fluctuations in spheres of $8\;h^{-1}Mpc$ of $\sigma_8=0.9$, and a $n=1$
slope for the primordial power spectrum. The simulation followed $2160^3$ particles
from $z=127$ to $z=0$ in a comoving periodic volume of $500\;h^{-1}Mpc$ a side. The resulting
galaxy population after applying the \citet{bow06} model can be considered complete down to an
absolute magnitude in the r-band of $M_r=-15$.
 
In order to check the presence of biases in the method that was used to compute the SL2S group masses (Section \ref{sec:mass}), it is necessary to first determine the observational selection and completeness effects. The dominant selection effect is the fraction of group members that were observed and classified as secure members of each SL2S group. \citet{cou09} computed the photometric redshifts for galaxies in the CFHTLS survey, and obtained a mean photometric redshift error of $\sigma_{z/(1+z)} \sim 0.038$ and an outlier rate of $\eta \sim 3\%$ using a sample of 1,532 galaxies (from W1 field) with secure spectroscopic redshifts. We estimated the total number of red-sequence galaxies for each group, $N_{gal}$, using a method similar to the one used by \citet{koe07} for building the MaxBCG cluster catalog, but we added photometric redshift information to reduce the contamination by background and foreground galaxies. We estimated $N_{gal}$ for each group by counting the number of galaxies within a radius of $1\;\rm{Mpc}$, having magnitudes brighter than $\rm{R}=22.5$ and colors within $|{g-R}|<0.24$ (equivalent to $2 \sigma_{\delta (g-R)}$) with respect to the E/S0 ridgeline, and having $|{z_{phot}-\bar{z}_{spec}}| \leq 0.038*(1+\bar{z}_{spec})$, i.e. $1 \sigma_z$. We found that the number of red-sequence galaxies in the SL2S groups goes between 2 and 64 galaxies (see Table \ref{tab:summary}). The fraction of group members with measured recessional velocities was estimated as the ratio between $N_{spec,RS}$ and $N_{gal}$, and it ranges from 0.25 to 0.85 (not considering the bimodal groups).

We select haloes from the $z=0.509$ simulation output and repeat as closely as
possible the observational procedure to measure cluster dynamical
masses. { The simulation cube consists of (X,Y,Z) spatial coordinates and ($v_\mathrm{X}$,$v_\mathrm{Y}$,$v_\mathrm{Z}$) velocities .
A first step consists of choosing the Z coordinate
axis in the simulation cube as the line of sight, and defining the recessional velocity by $\mathrm{Z}\times100\;\mathrm{h\;km\,s^{-1}\,Mpc^{-1}}+v_{\mathrm{Z}}$.}
Since all the groups studied in this work are bona fide
gravitational lensing systems,
we assume the sample to be free from spurious groups and clusters, and therefore use the full sample of haloes to do
these tests.

\begin{figure}
\centering
\resizebox{\hsize}{!}{\includegraphics{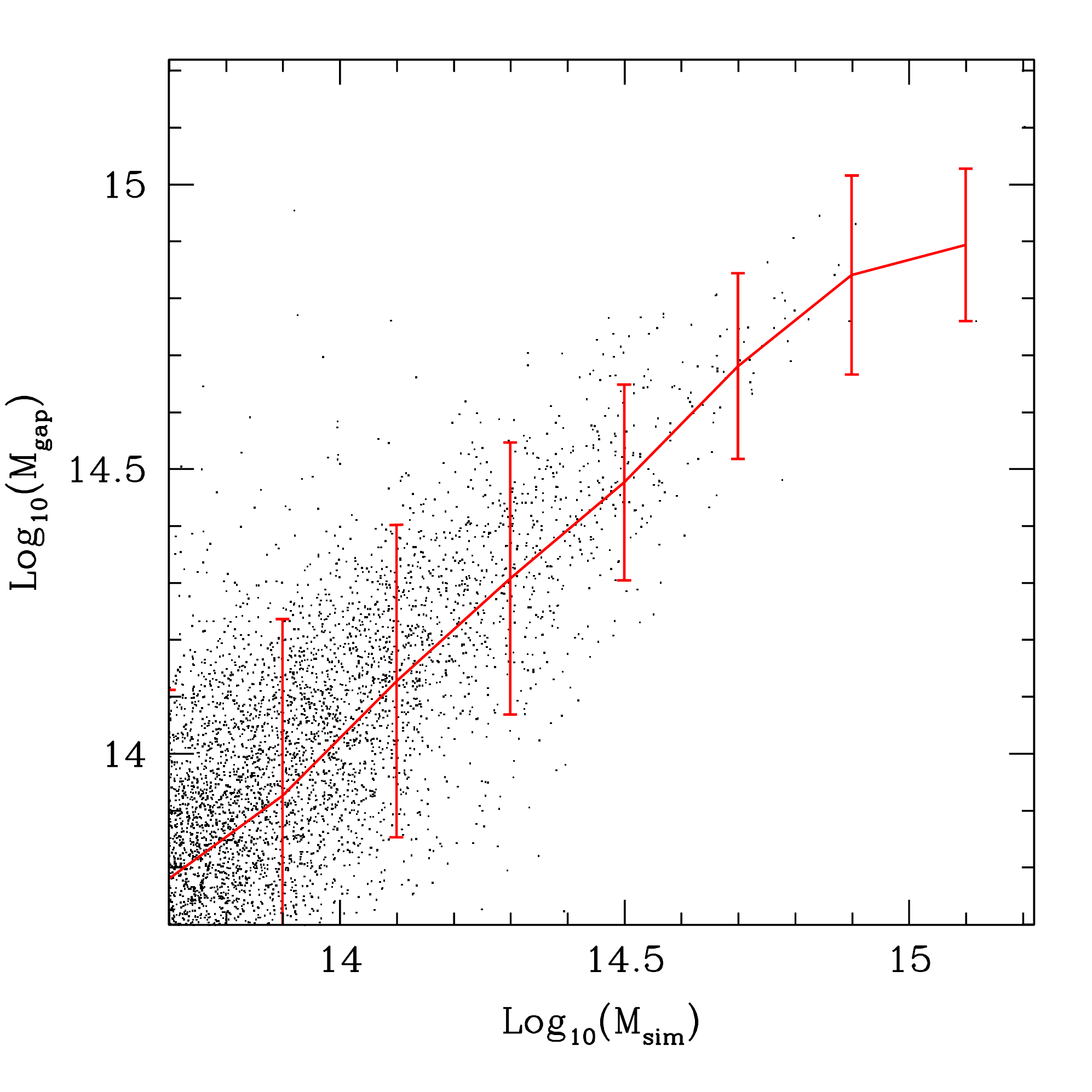}}
\caption{Comparison between estimated gapper mass and underlying halo mass in the simulation.
{ This shows the results when using a 30\% of the group members brighter than $M_r=-18$, randomly selected.} Dots represent individual
measurements; the solid lines correspond to the median in bins of halo mass, and the errorbars
enclose $68$ percent of the individual measurements.}
\label{fig:masses}
\end{figure}

For each individual dark-matter halo we select galaxies in the red sequence (defined using
empirical color cuts) in a cylinder with depth $\Delta v=500\;km\,s^{-1}$
and width $\Delta \theta=220\arcsec$, transformed into comoving coordinates at the redshift
corresponding to the selected output ($z=0.509$).  {These values of $\Delta v$ and $\Delta \theta$ are
iteratively corrected once the velocity dispersion and harmonic projected radius of the halo are
obtained from the possible members of the halo.}  Their final values are used to calculate the
gapper mass of the halo. Figure \ref{fig:masses} shows a comparison between the recovered and simulated
masses of group-size dark matter haloes (the latter being simply the number of dark-matter particles per halo multiplied by the particle mass), where
it can be seen that when the member galaxies are those
brighter than $M_r=-18$ { (which corresponds to the observed $i=22\;\mathrm{mag}$ for a group at $z\sim0.4$)}, the gapper method introduces important
uncertainties in the recovered masses of about of $20\%$ and $50\%$, for
masses of $\sim 10^{14}h^{-1}M_{\odot}$ and $\sim 10^{15}h^{-1}M_{\odot}$ respectively;
notice the significantly larger values compared to the results by \citet{biv06}, mainly
due to the low number of members available in our observational samples.

\begin{figure*}
\centering
\begin{minipage}{9 cm}
\includegraphics[width=8.5cm]{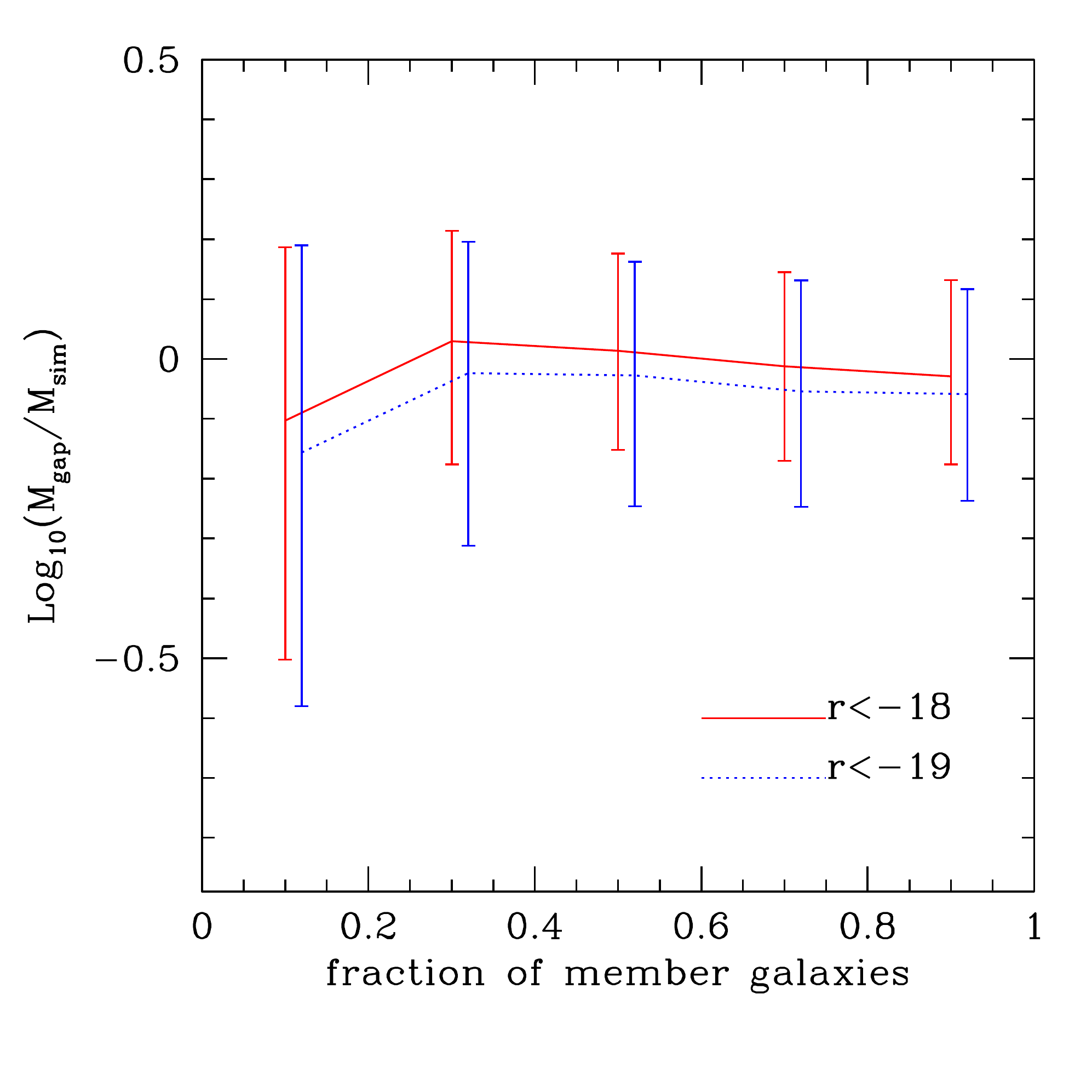}\\
\end{minipage}
\hfill
\begin{minipage}{9 cm}
\includegraphics[width=8.5cm]{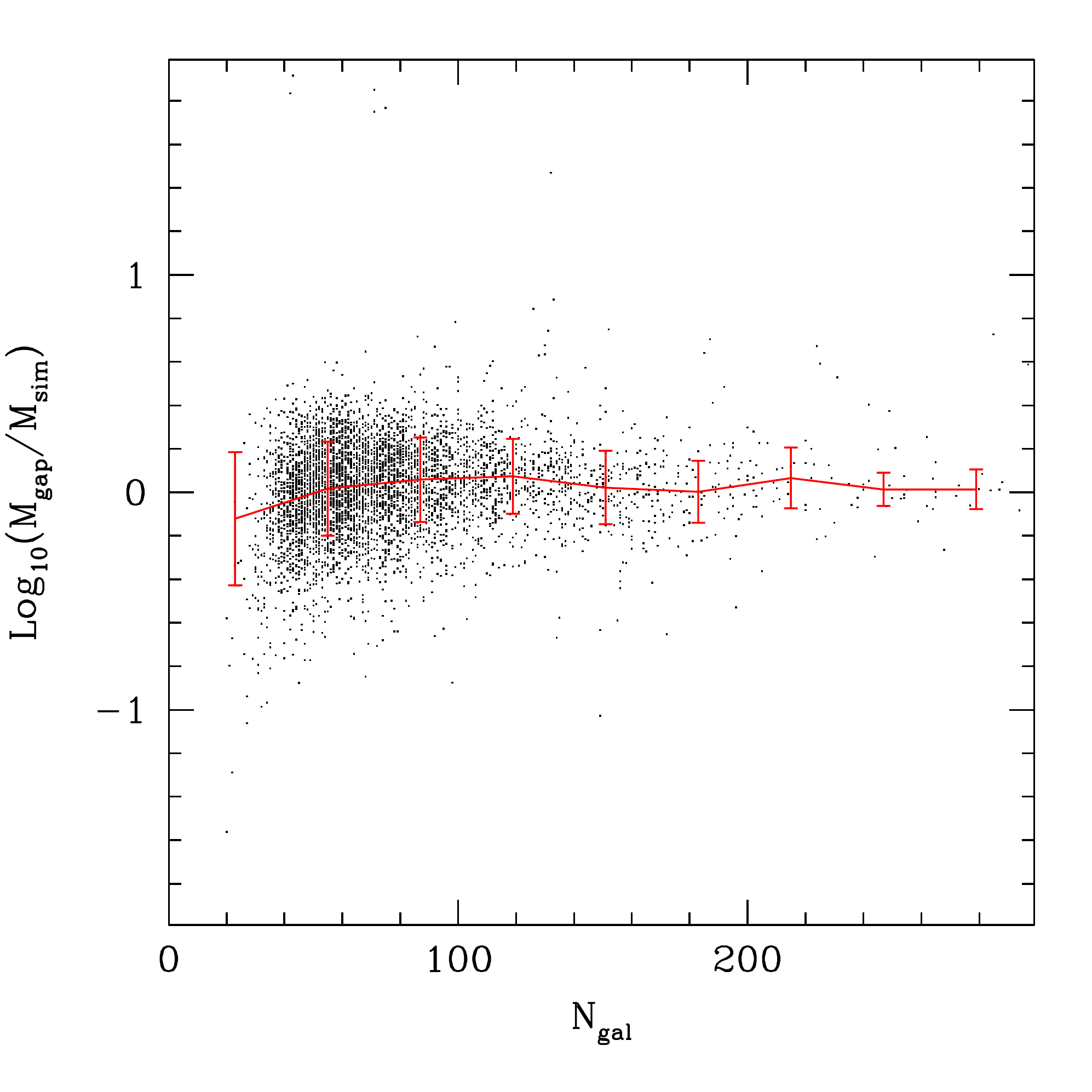}\\
\end{minipage}
\hfill
\begin{minipage}{9 cm}
\includegraphics[width=8.5cm]{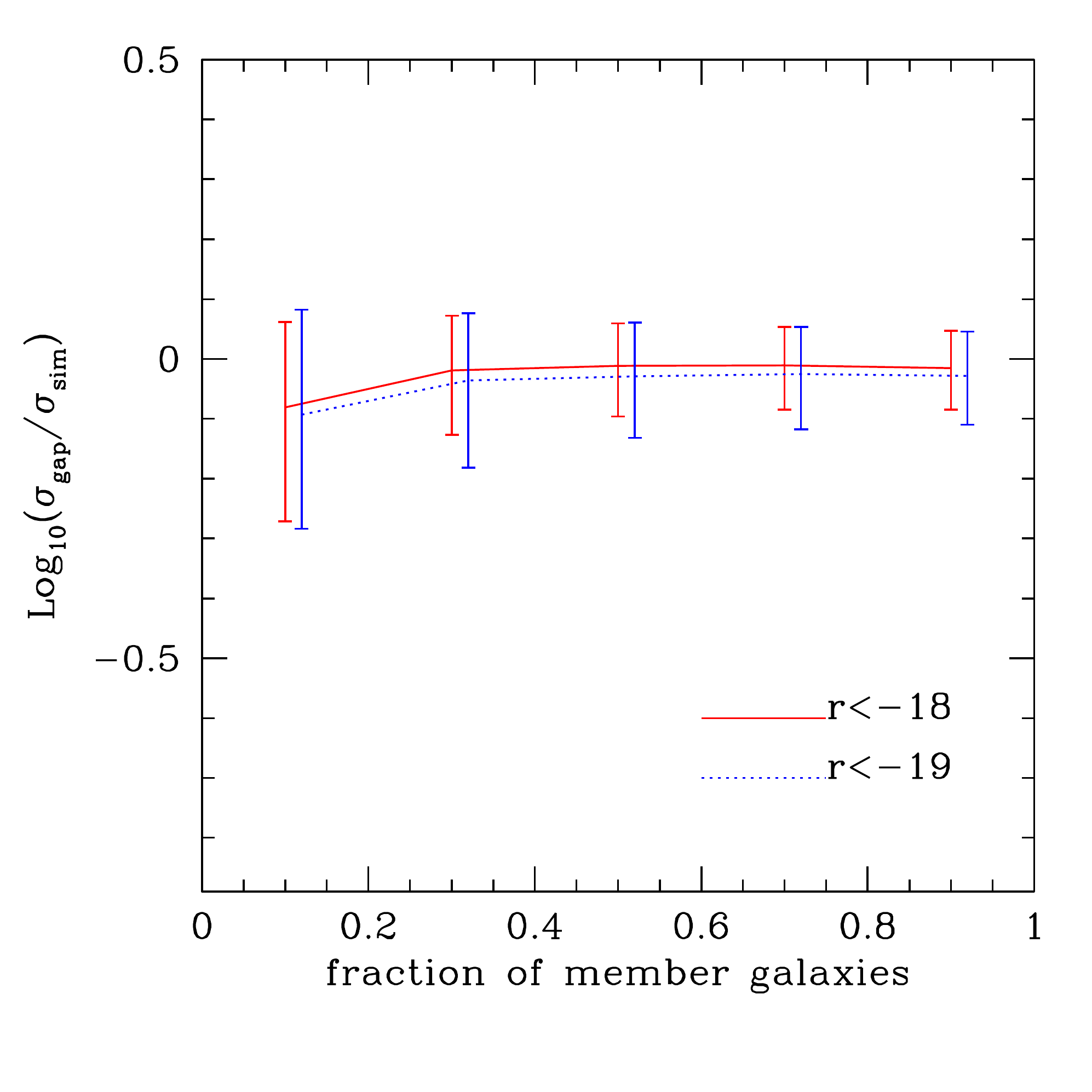}\\
\end{minipage}
\begin{minipage}{9 cm}
\includegraphics[width=8.5cm]{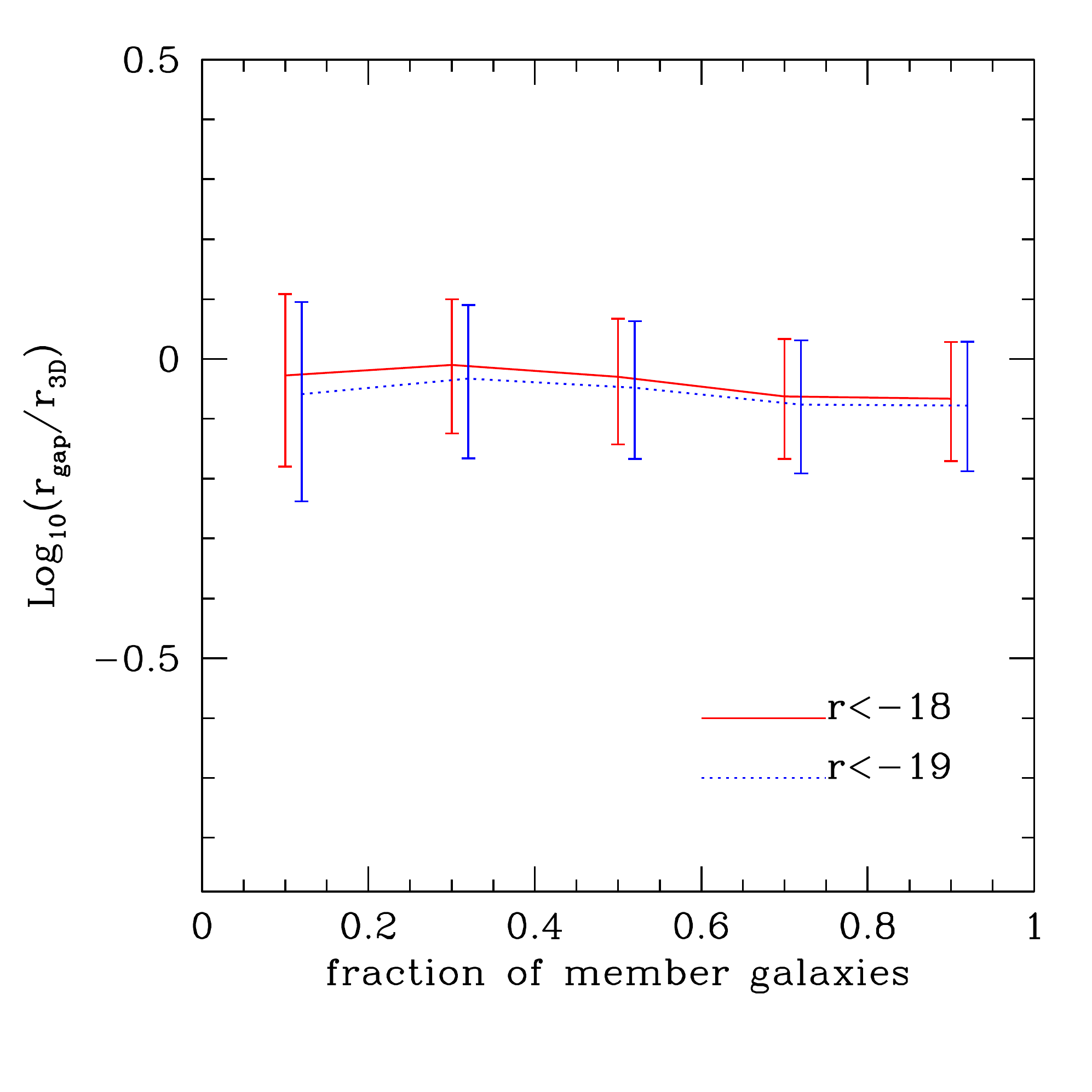}
\end{minipage}
\caption{\textit{Top left:} Ratio between the estimated and actual masses of group-size dark matter haloes as a function
of the fraction of group members with recessional velocity measurements, as obtained from the numerical simulations
presented in this work.
\textit{Top right:}  Ratio between the estimated and actual masses of group-size dark matter haloes as a function
of the number of member galaxies.
\textit{Bottom left:} Ratio between the velocity dispersion as estimated with the gapper method and as obtained from the actual group members
in the simulation.
\textit{Bottom right:} Ratio between the deprojected virial radius and the mean 3D radius of actual groups members in the simulation,
as a function of the fraction of members.
The solid red lines show the results when using only members with absolute magnitudes brighter than $M_r=-18$,
and the dotted blue lines for members brighter than $M_r=-19$. Errorbars enclose 68\% of the measurements in each bin of fraction.}
\label{fig:ratios}
\end{figure*}

A more detailed interpretation of the simulation tests is shown in Figure \ref{fig:ratios}.  The top left panel
shows the ratio between the gapper and simulated mass of the dark matter haloes, as function of the number
of group members that were selected to estimate their respective group virial masses. { We use the entire sample of galaxy groups from the simulation, and on average the number of members is 60.} As can be seen, the statistical
errors shown by the error bars (enclosing $68$ percent of the individual results) is significantly larger
than the systematic bias.  In the case of the larger sample of galaxies (selected with $M_r<-18$),
the bias is almost null, with a slight tendency of recovering a lower value for the estimated
mass as the fraction increases.  The same conclusion applies for the analysis using the brighter sample of members, with
the difference that regardless the fraction of members used in the analysis, the estimated group masses are always lower than their
simulated counterparts. It is important to note that biases in the mass estimate are lower than $20$ percent for all the cases.
Also, we note that there is an important drop in the error bars as soon as the fraction of group members used in the calculation
of the mass increases above $20\%$. { It is important to notice that the average number of true members is constant across the x-axis, i.e. we only change the number  of spectroscopic members used to estimate the group mass.  The top right panel shows the mass ratio as a function of the total number of members, which spans between 20 and 300 members on average.} 
As discussed in Section \ref{sec:obs}, the mean fraction of group members with spectroscopic measurements for the SL2S groups is
about $30\%$; as can be seen, using a larger fraction would not have resulted in much of an improvement (and would require significantly
more telescope time).

In order to account for the sources of uncertainty in the estimation of the virial mass, { in the bottom left panel of Figure
\ref{fig:ratios}} we show the ratio between the group velocity dispersions estimated by using the gapper method and
the ones obtained from the full numerical simulation, and in the { bottom right panel} we show the ratio between the estimated
harmonic radii, $r_{gap}$, and the average three-dimensional radii of simulated dark matter haloes, $r_{3D}$. Our results suggest that the velocity
dispersion is always underestimated, and that it gets closer to the underlying value as the fraction of observed group members increases
(with little change above a fraction of $0.3$). { Furthermore, we found that the virial radius is also underestimated and it
departs from the actual value as the fraction of members increases. We note that the ratio $r_{gap}/r_{3D}$ can not be compared with the ratios in columns 9 and 10 of Table 1 from \citet{biv06}, since we use a different method to estimate the virial radius.}


\begin{figure}[htp]
\centering
\resizebox{\hsize}{!}{\includegraphics{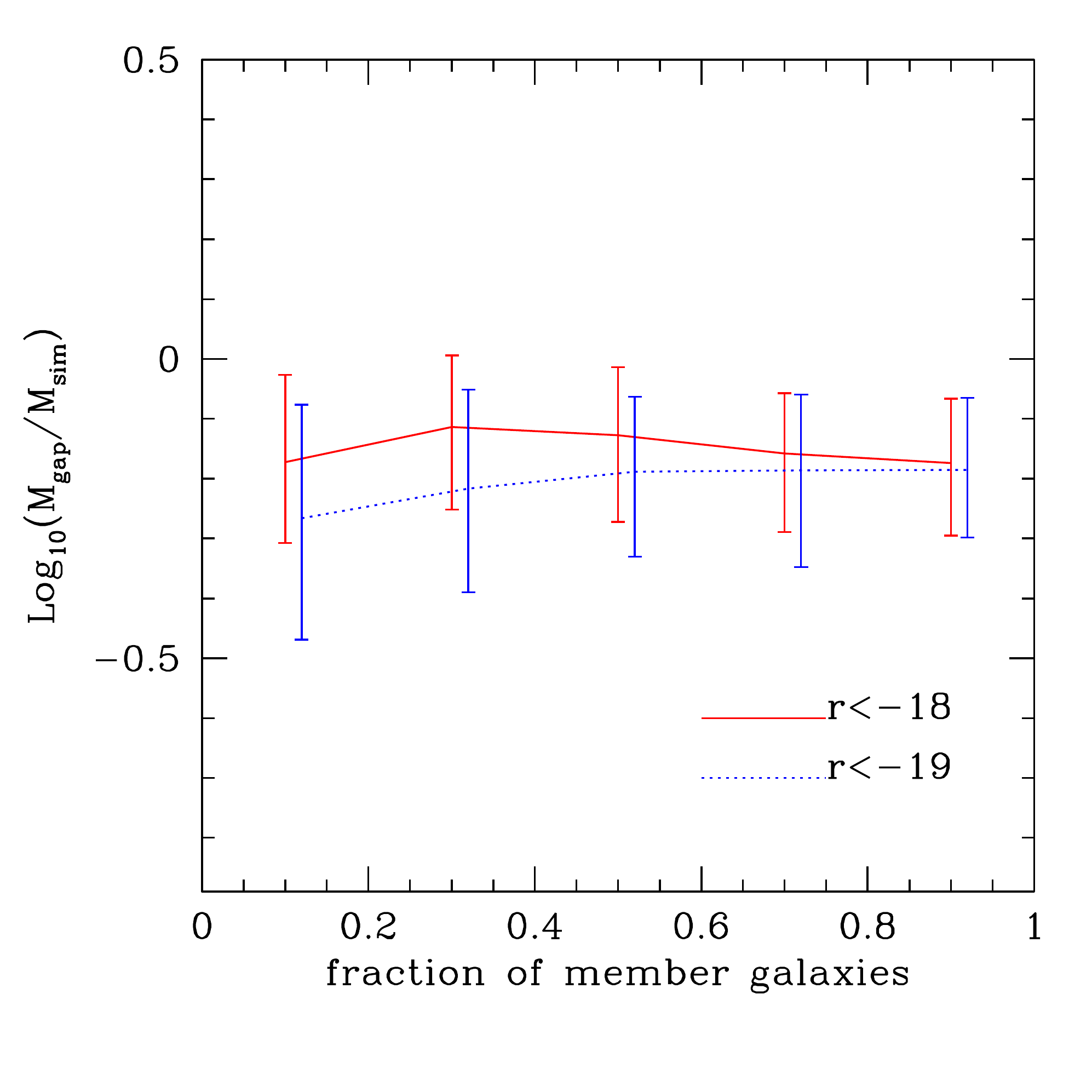}}
\caption{Ratio between the estimated and actual masses of simulated groups with total number of members
and velocity dispersions similar to the ones measured for \mbox{SL2SJ02140-0535}, i.e. number of members between 20 and 40, and velocity dispersion between 260 and $460\;km\,s^{-1}$. The number of groups in each bin of the plot remains constant. The x-axis corresponds to the fraction of members used for the calculation of the mass.}
\label{fig:indiv}
\end{figure}

The previous tests were applied to bound systems with masses larger than $10^{14}\;M\odot$, but could also be applied to a subset
of simulated groups of masses similar to any of the observed SL2S groups.
We chose the group \mbox{SL2SJ02140-0535} as a particular case for applying the test, since this group has the largest number of confirmed
members and a virial mass properly covered by the numerical simulation \citep{ver11}.
We selected haloes from the numerical simulation with velocity dispersions
within a narrow range around the measured value for \mbox{SL2SJ02140-0535} ($264<\sigma(v)_{los}<464\;km\,s^{-1}$), and restricted to haloes having a total number of $M_r<-21$
galaxies between $20$ and $40$, which brackets the estimated number of members for this group.
The main aim of studying this restricted sample of groups was to obtain the systematic
and statistical uncertainties in the estimated mass of \mbox{SL2SJ02140-0535}. The ratio between the estimated and actual
masses for this subsample of haloes is shown in Figure \ref{fig:indiv}, where as can be seen,
when using galaxies brighter than $M_r=-18$ the resulting biases are comparable to those
found for the full sample of haloes in the numerical simulation.

The main conclusions are that the expected biases in the
virial mass estimation are lower than the statistical uncertainties of group properties in a narrow range of mass,
and that the virial mass is underestimated by about $\sim15$ percent when using $30\%$ of the group members.
According to Figure \ref{fig:indiv}, the bias appears to worsen for higher fractions of members, but this effect
is likely due to the small sample of haloes resulting from the cuts applied to mimic \mbox{SL2SJ02140-0535}.

\section{Discussion}
\label{sec:discussion}

The formation and evolution history of galaxies has been shown to be strongly dependent on the properties of the environment they inhabit. The existence of this environmental dependance has been confirmed both in the local \citep{got03,hel03,bam09} and intermediate redshift Universe \citep{dre97,tre03,pan09}, and the observational results suggest that star formation and galaxy merging processes are accelerated in high density environments such as galaxy groups and clusters.

Therefore, galaxy groups represent natural laboratories to study the relative importance of the different astrophysical processes occurring in dense regions of the Universe. For instance, galaxy collisions are expected to be most effective in groups because the dynamical friction timescale is similar to the orbital timescale of galaxies within the group. Since galaxy groups contain a low number of members (less than hundred within $1\; \rm{Mpc}$) and cover a wide range in mass, velocity dispersion and hot gas content, it is necessary to characterize their properties in detail in order to build representative samples of groups and to properly study the relative importance of the different mechanisms.

The virial mass estimation relies on the following assumptions: sphericity, kinematic isotropy and virialization. In our virial mass estimation we use the velocity dispersion, which we assume constant through the group, and the harmonic radius as an estimate of the virial radius (see Section \ref{sec:redshift}). These assumptions depart slightly from observational results. For instance, i) groups and clusters usually show substructure \citep{rie09}, ii) it is known that nearby clusters show a small gradient in the velocity dispersion \citep{ken82}, iii) some groups at high redshift show evidence of having  merging events, and therefore are not in a virialized state \citep{mck10}. Furthermore, it is a well known fact that the harmonic radius depends strongly on the area covered by the spectroscopic observations and the number of confirmed members \citep{biv06,gir98}, biasing the measurement of the virial radius. {From our simulations, it seems that the virial radius is underestimated when the number of members increases, contrary to the results obtained by \citet{biv06}. The difference between our results and those from \citet{biv06} could be explained due to differences in the simulations, since the latter uses a N-body hydrodynamical simulation.}

We ran numerical simulations to assess the bias introduced in the virial mass estimation (see Section \ref{sec:sim}). The results show that the method adopted in this work allows to recover the mass of the simulated groups within the error bars, showing no systematic deviations (see top panel of Figure \ref{fig:ratios}). It is also important to note that: i) the mass estimation of a group with 40\% of their members observed is as good as when using 90\% of the members, ii) the mass estimation improves with the number of groups used, and justifies the stacking of groups of similar properties in order to obtain better estimations. However, for the particular case in which we applied a cut in velocity dispersion, the case of SL2SJ02140-0535, the simulation shows that the method underestimates the group mass by about 20\% (see Figure \ref{fig:indiv}). The analysis of simulated groups selected from the numerical simulations (see Section \ref{sec:sim} and Figure \ref{fig:ratios}) reveals that the virial mass of the SL2S groups is underestimated by 15\%.



\citet{lim09} computed the weak lensing velocity dispersions and masses for several of the SL2S systems studied in this work. The mass estimate obtained from weak lensing measurements makes no assumption regarding the dynamic state of the systems, as opposed to the kinematical measurements presented in this work. Although weak lensing requires less assumptions, it has been shown that the estimated mass of galaxy clusters can be strongly affected by intervening large-scale structure along its line of sight \citep{hoe11}.

\begin{figure}
\centering
\resizebox{\hsize}{!}{\includegraphics{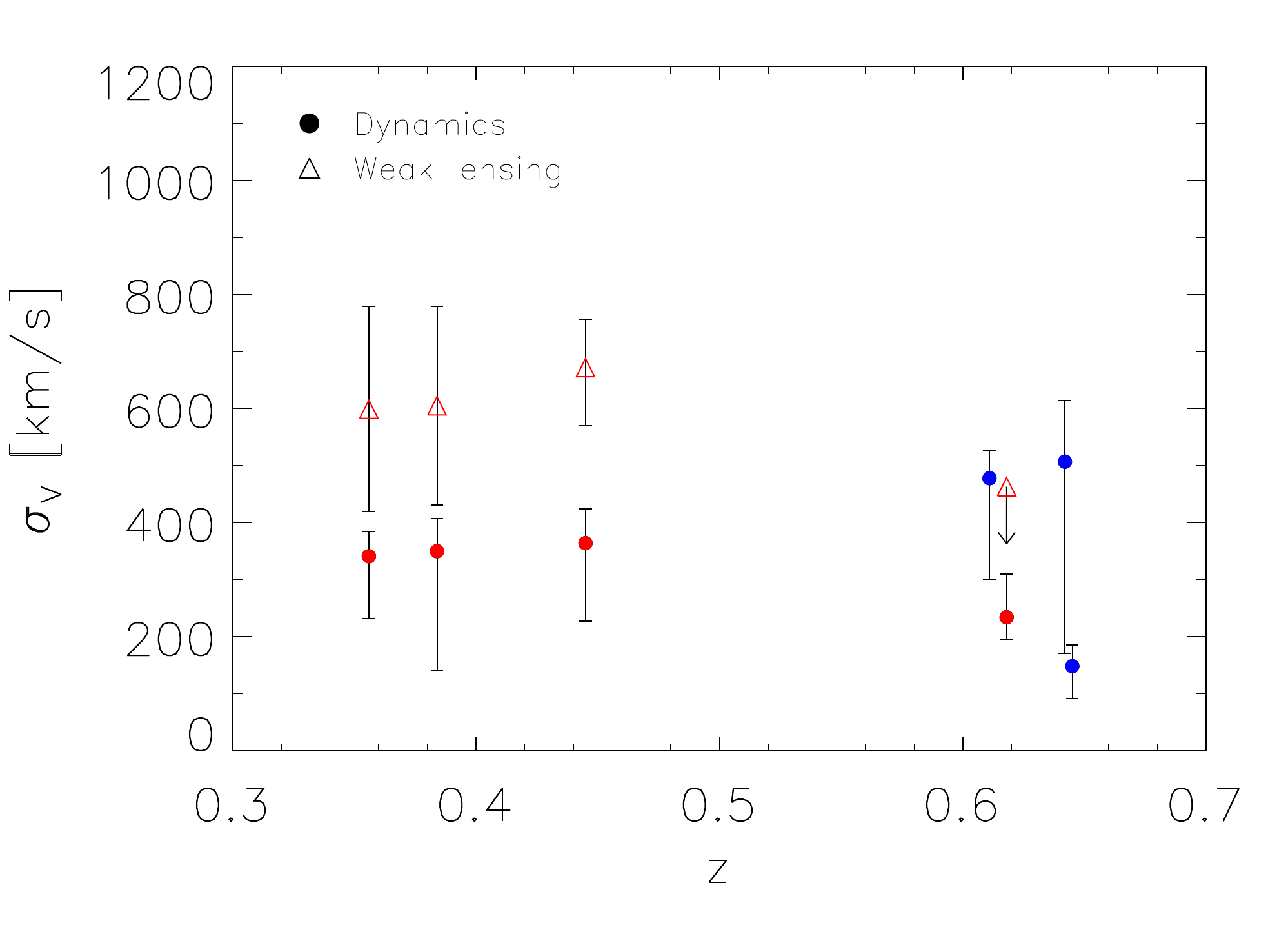}}
\caption{Velocity dispersion of SL2S groups as estimated from galaxy dynamics (solid circles) and weak lensing measurements (open triangles). The red symbols show the groups with dynamical and weak lensing estimates, while blue symbols those groups with only dynamical estimates. Only one group at $z>0.6$ has enough weak lensing signal to estimate an upper limit to its velocity dispersion.}
\label{fig:velocity_comparison}
\end{figure}

{Figure \ref{fig:velocity_comparison} shows the line-of-sight velocity dispersions as estimated from dynamics and weak lensing measurements for the groups \mbox{SL2SJ02140-0535}, \mbox{SL2SJ02215-0647}, \mbox{SL2SJ08544-0121} and \mbox{SL2SJ09413-1100}. We found that weak lensing estimates are systematically larger by 50\% than dynamical estimates. This is in stark contrast with measurements from galaxy-galaxy lensing by SLACS \citep{treu06a,treu06b}, where the ratio between the central stellar velocity dispersion and the velocity dispersion that best fit the lensing model is $1.01\pm0.02$. Although numerical simulations (see Section \ref{sec:sim}) suggest that the viral mass is underestimated by 20\%, this is not enough to explain the inferred discrepancy. Thus, our results may indicate that the isothermal sphere model is not a good assumption for galaxy groups.}

\section{Conclusions}
\label{sec:conclusions}

This paper presents the spectroscopic follow-up and dynamical analysis of 8 group candidates identified in the SL2S survey. Our analysis reveal that 7 of the systems correspond to gravitationally bound structures, where 5 have a single component in redshift-space and 2 have a double component (bimodal galaxy distribution). They span a wide range in redshift between 0.35 and 0.65, and a wide range in mass between $5\times 10^{13}\;M_{\odot}$ and $1.5\times 10^{14}\;M_{\odot}$.

The main results of this paper are given as follows,

\begin{enumerate}
\item The success rate of the spectroscopic confirmation of groups identified in the SL2S survey is about 88\%. It is similar to the success rate of clusters followed-up in the Red-Sequence Cluster Survey \citep[RCS-1;][]{gla05}.
\item { We found that weak lensing estimates of the group velocity dispersions are 50\% larger than dynamical estimates. This discrepancy has been never reported before by other studies in groups and clusters, and is in stark contrast with measurements from galaxy-galaxy lensing.}
\item From numerical simulations, we conclude that measuring redshifts for only 30\% of the total galaxy population in groups is enough to recover the group velocity dispersion with less than 5\% systematic error.
\end{enumerate}

\begin{acknowledgements}

We thank Ricardo Salinas and Anupreeta More for enlightening conversations. We thank the referee for their useful and constructive comments to improve this paper.\\
The authors acknowledge support from Fondo ALMA-CONICYT $\rm{N}^\circ\,31090019$, Comit\'e Mixto ESO-Gobierno de Chile and FONDECYT through grant 1090637.\\
R.P. Mu\~noz acknowledges support from CONICYT CATA-BASAL and FONDECYT through grant 3130750.\\
V. Motta gratefully acknowledges support from FONDECYT through grant 1090673 and 1120741.\\
T. Verdugo acknowledges support from FONDECYT through grant 3090025 and CONACYT through grant 165365.\\
M. Limousin acknowledges the Centre National de la Recherche Scientifique for its support and the Dark Cosmology Centre is funded by the Danish National Research Foundation.\\
N. Padilla acknowledges support from FONDAP ``Center for Astrophysics'', CONICYT CATA-BASAL and FONDECYT through grant 1110328.\\
G. Fo\"ex acknowledges support from FONDECYT through grant 3120160.\\
R. Gavazzi acknowledges support from the Centre National des Estudes Spatiales.\\ \\
This work made use of the Geryon cluster at the Centro de Astro-Ingenier\'ia UC.
\end{acknowledgements}

\bibliographystyle{aa} 
\bibliography{bibs} 
\clearpage

\Online
\onecolumn

\begin{appendix}
\section{Presentation of each group}
\label{sec:ap1}

\begin{figure*}[!htp]
\centering
\includegraphics[width=0.8\textwidth]{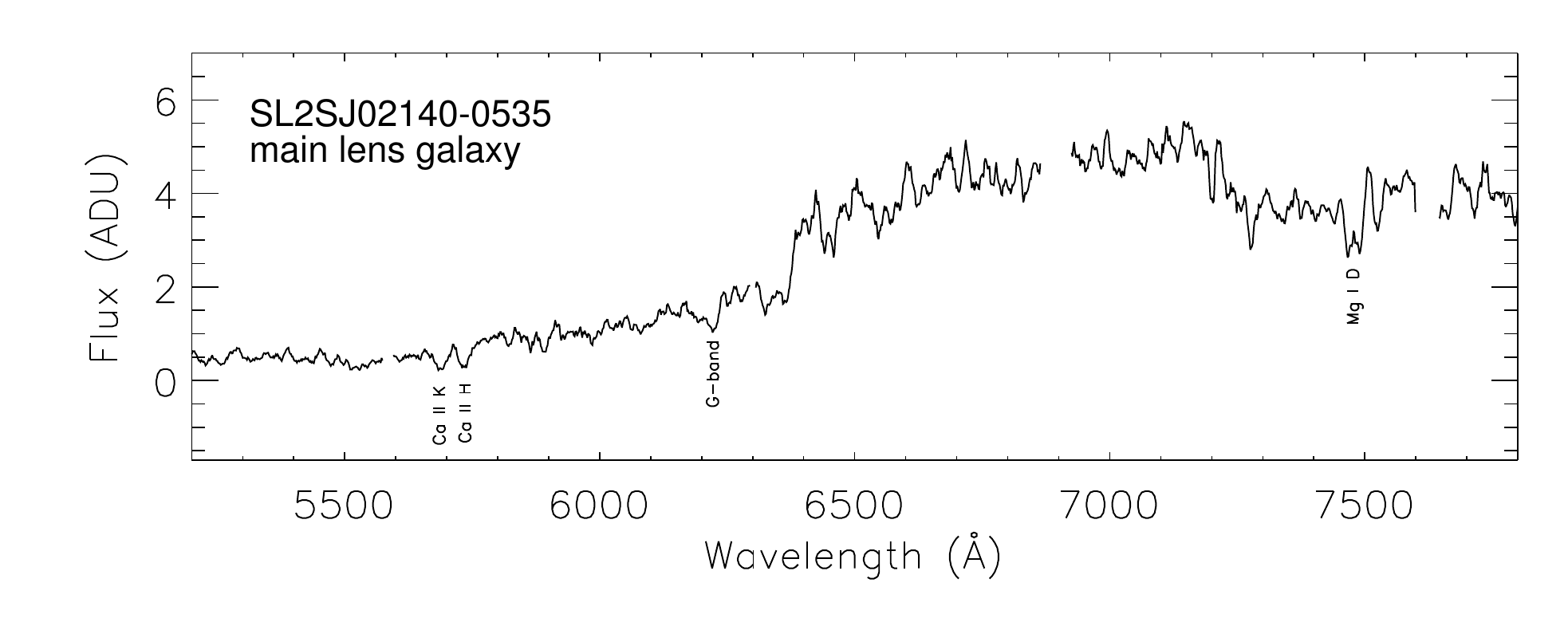}
\includegraphics[width=0.8\textwidth]{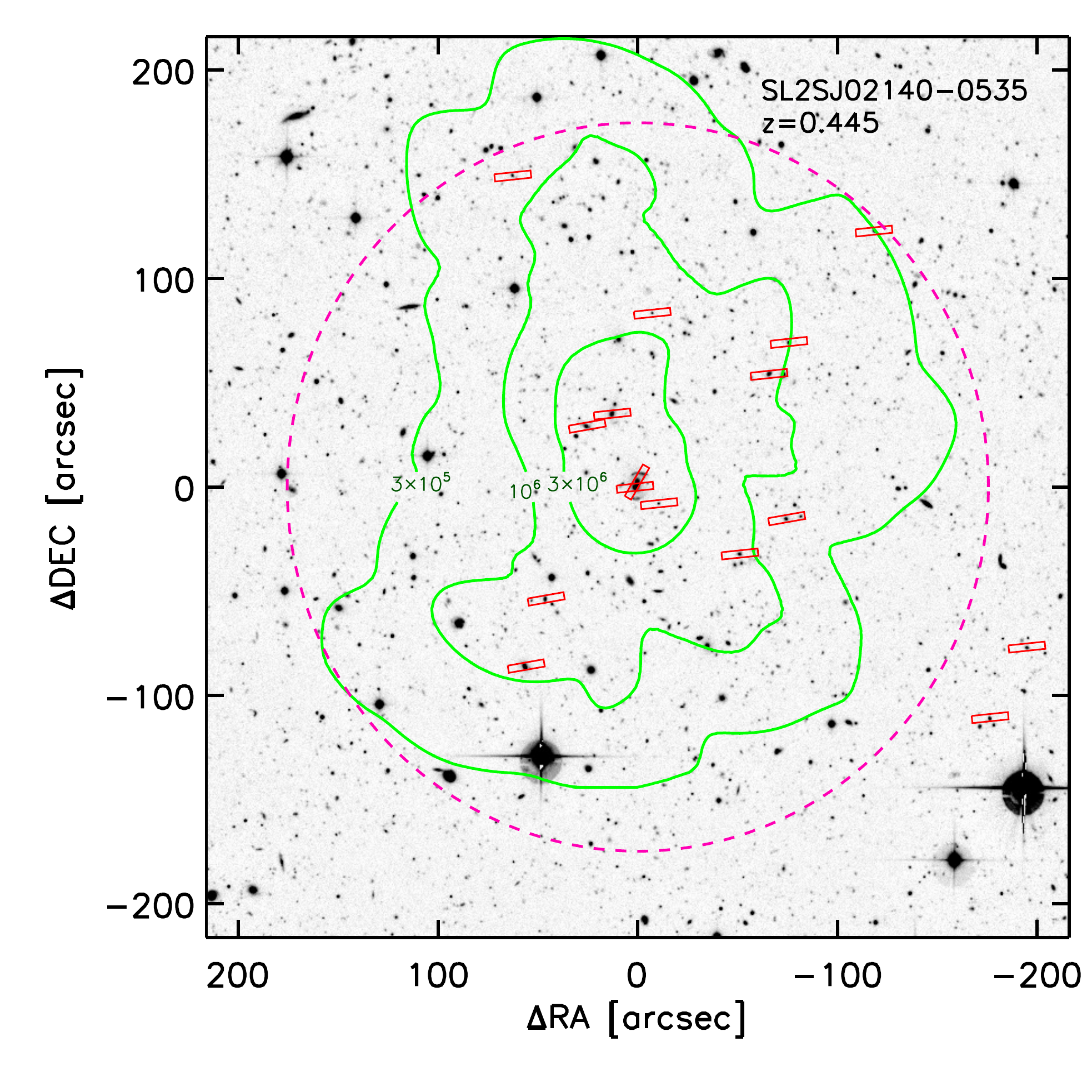}
\caption{Top panel: Optical spectra of the brightest confirmed member of SL2SJ02140-0535. The main absorption lines used to determine the redshift of the main lens galaxy have been identified.
Bottom panel: Spatial distribution of galaxies in the group field of SL2SJ02140-0535. Red rectangles show the position of the spectroscopically confirmed members. The dashed magenta circle shows a circular aperture of radius $1\;Mpc$ at z=0.44. The contours in green show the luminosity contours equal to $3\times10^5$, $10^6$, $3\times10^6$ and $10^7\;L_\odot\,kpc^{-2}$ from outermost to innermost, as computed by \citet{lim09}.}
\label{fig:spatial1}
\end{figure*}

\begin{figure*}[!htp]
\centering
\includegraphics[width=0.8\textwidth]{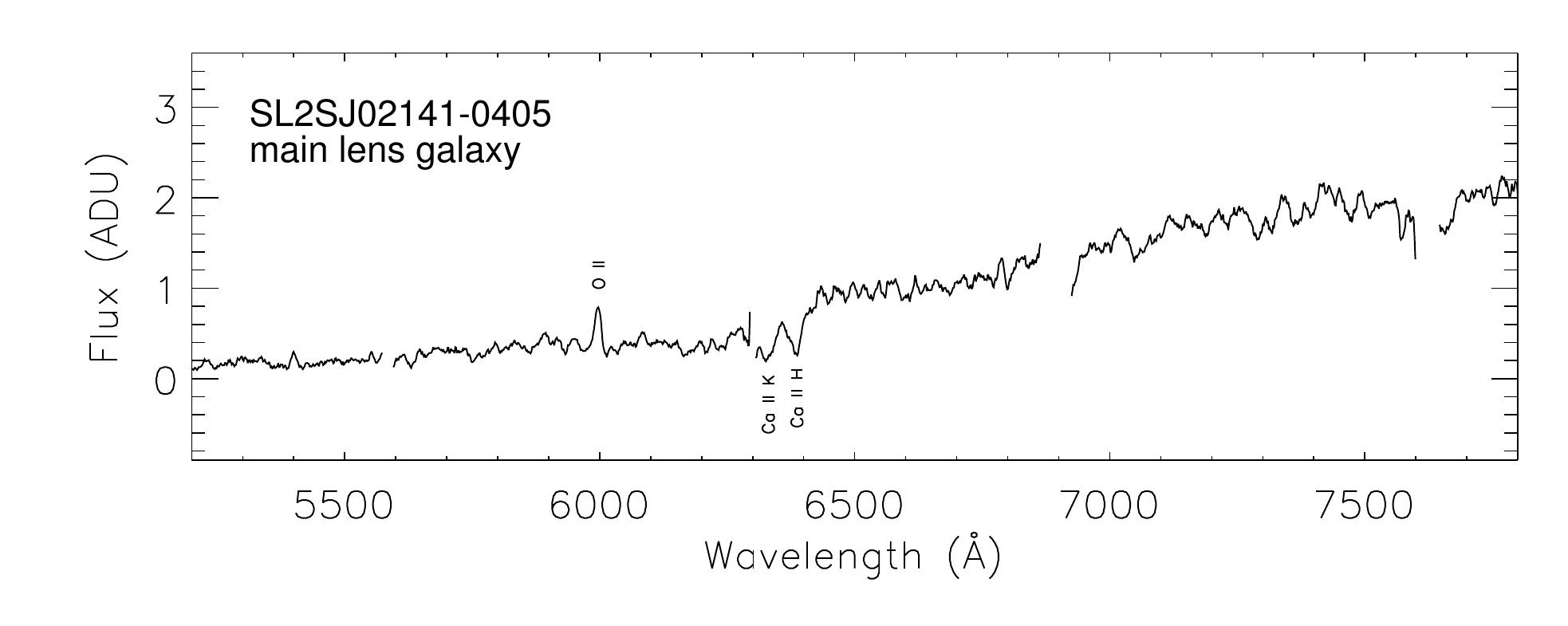}
\includegraphics[width=0.8\textwidth]{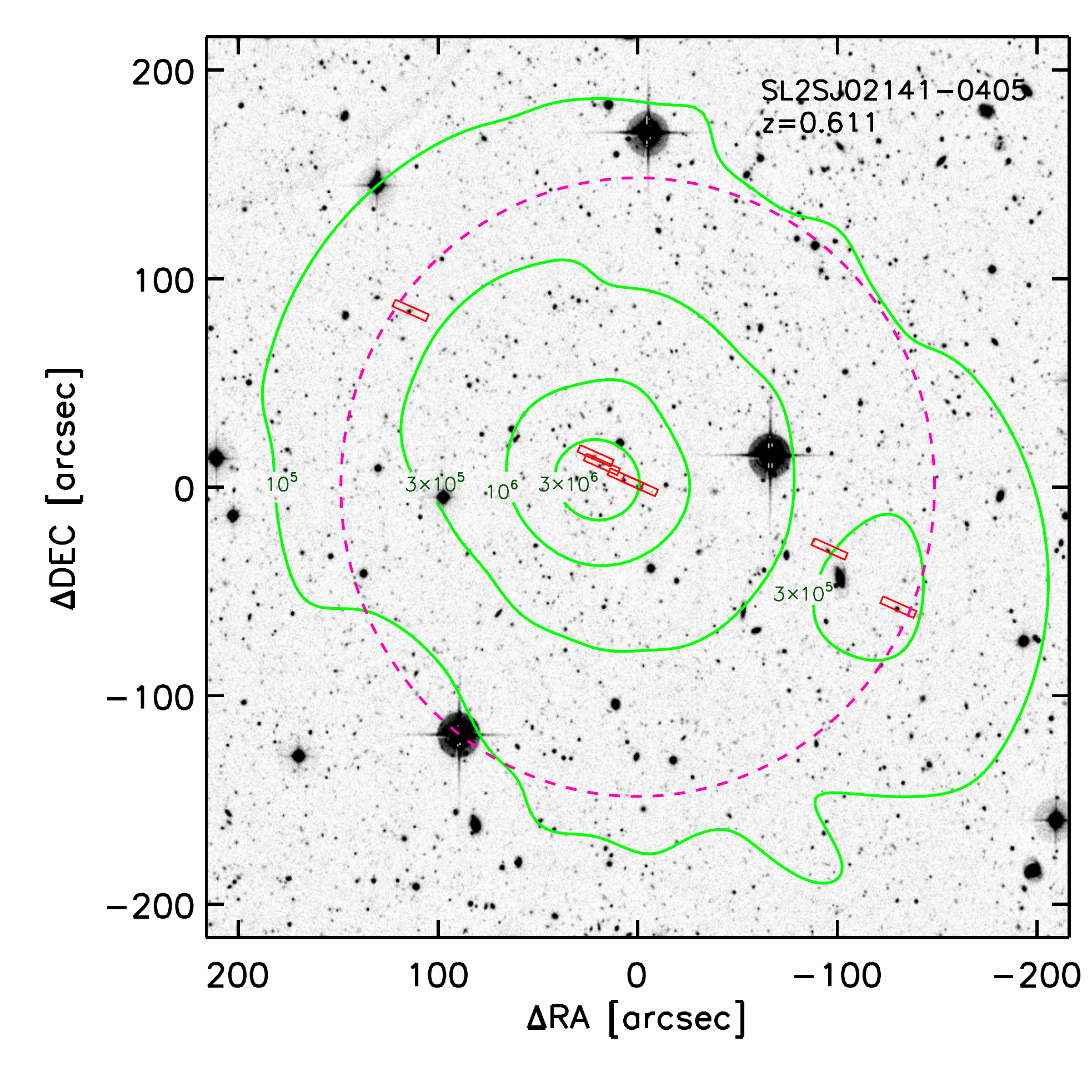}
\caption{Top panel: Optical spectra of the brightest confirmed member of SL2SJ02141-0405. The main absorption and emission lines used to determine the redshift of the main lens galaxy have been identified.
Bottom panel: Spatial distribution of galaxies in the group field of SL2SJ02141-0405. Red rectangles show the position of the spectroscopically confirmed members. The dashed magenta circle shows a circular aperture of radius $1\;Mpc$ at z=0.61. The contours are the same as in Fig. \ref{fig:spatial1}.}
\label{fig:spatial2}
\end{figure*}

\begin{figure*}[!htp]
\centering
\includegraphics[width=0.8\textwidth]{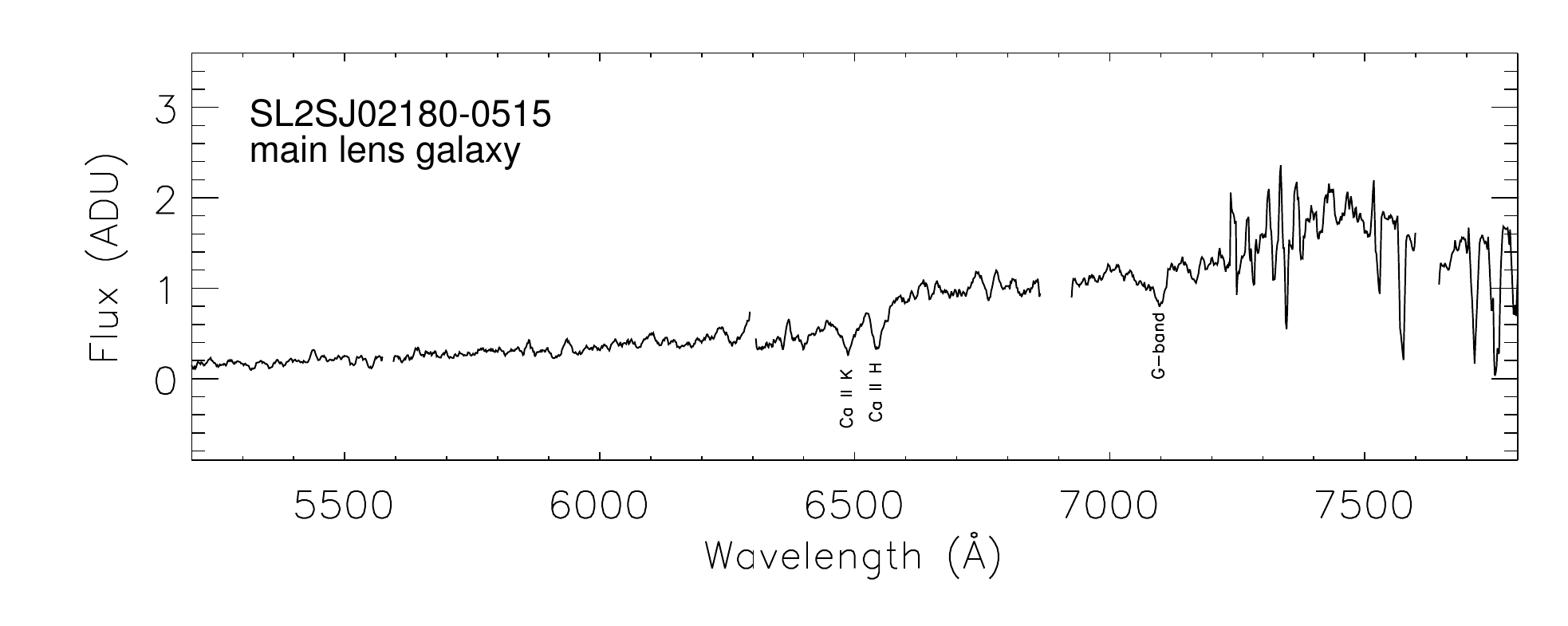}
\includegraphics[width=0.8\textwidth]{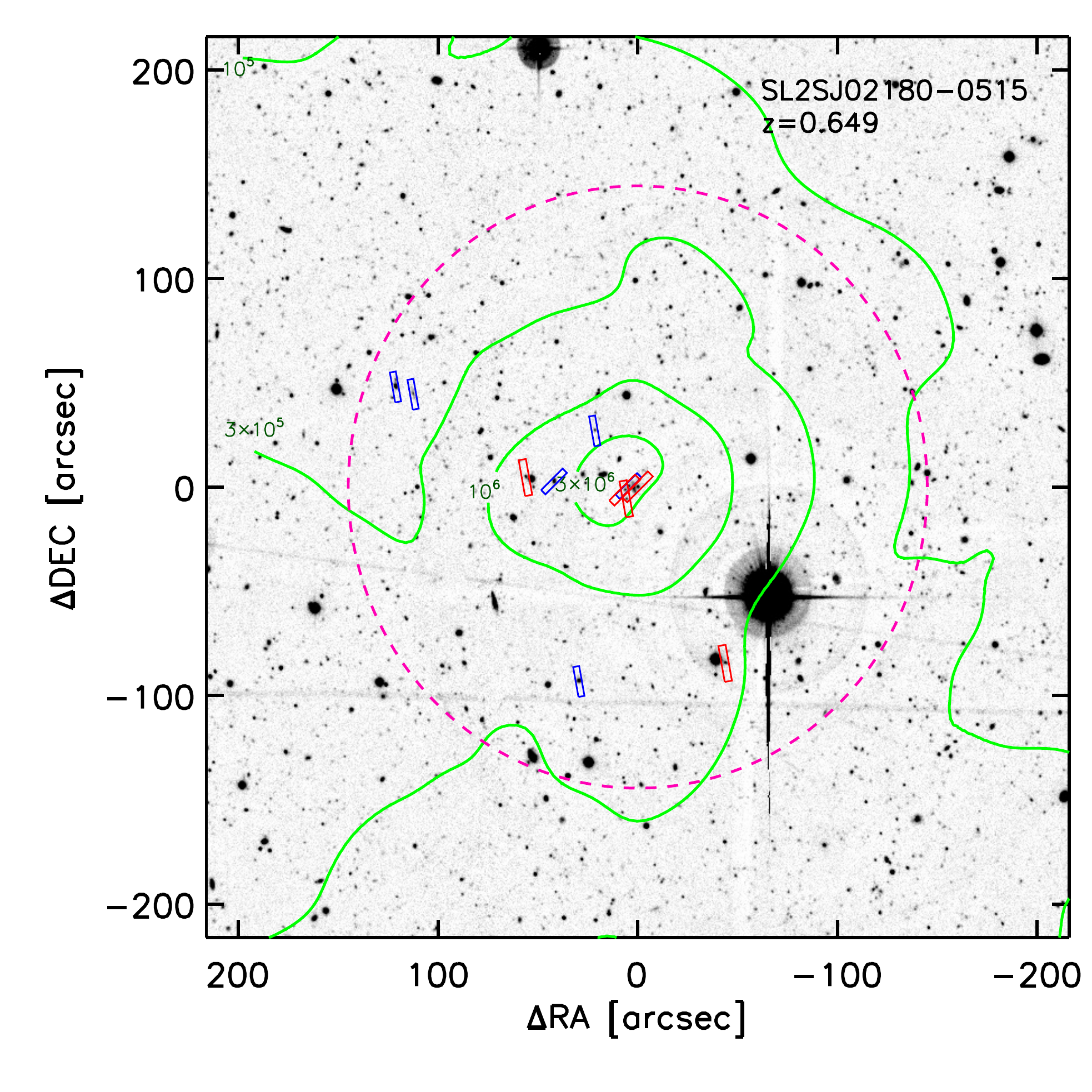}
\caption{Top panel: Optical spectra of the brightest confirmed member of SL2SJ02180-0515. The main absorption lines used to determine the redshift of the main lens galaxy have been identified.
Bottom panel: Spatial distribution of galaxies in the group field of SL2SJ02180-0515. Red rectangles show the position of the spectroscopically confirmed members. The dashed magenta circle shows a circular aperture of radius $1\;Mpc$ at z=0.64. The contours are the same as in Fig. \ref{fig:spatial1}.}
\label{fig:spatial2}
\end{figure*}

\begin{figure*}[!htp]
\centering
\includegraphics[width=0.8\textwidth]{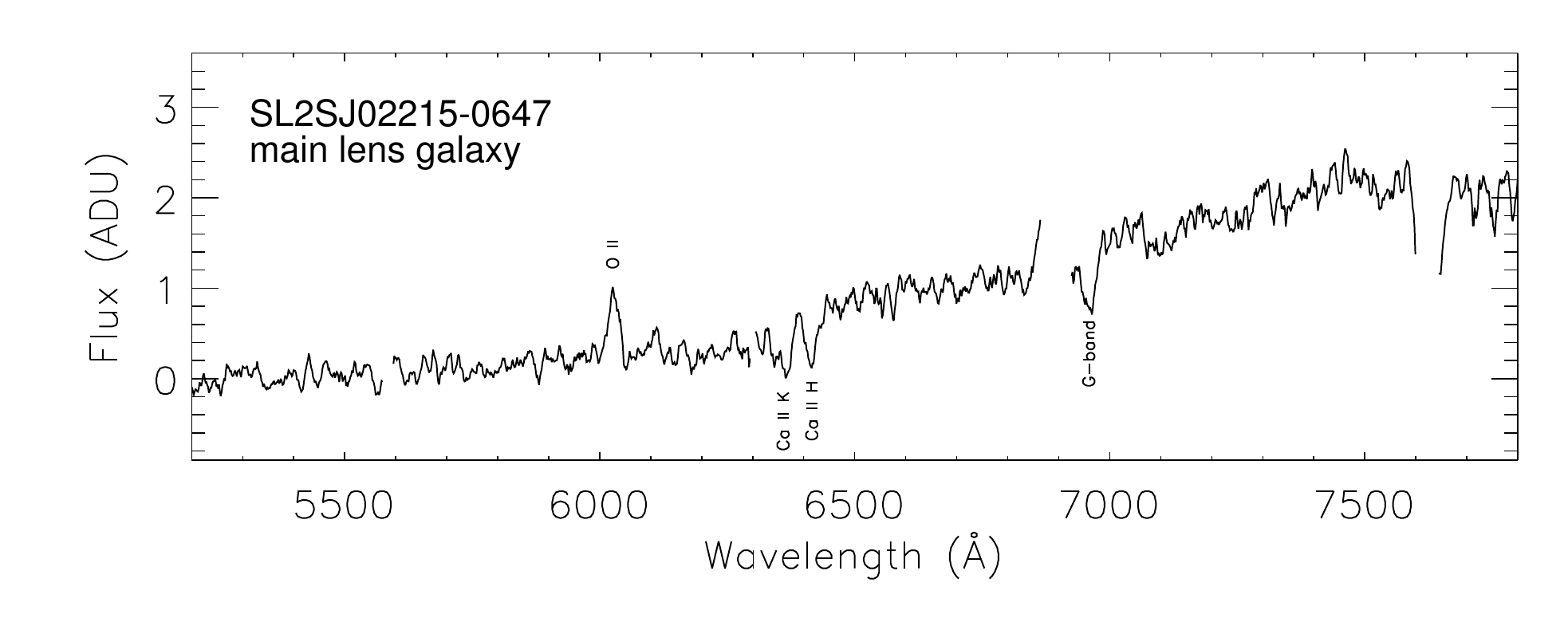}
\includegraphics[width=0.8\textwidth]{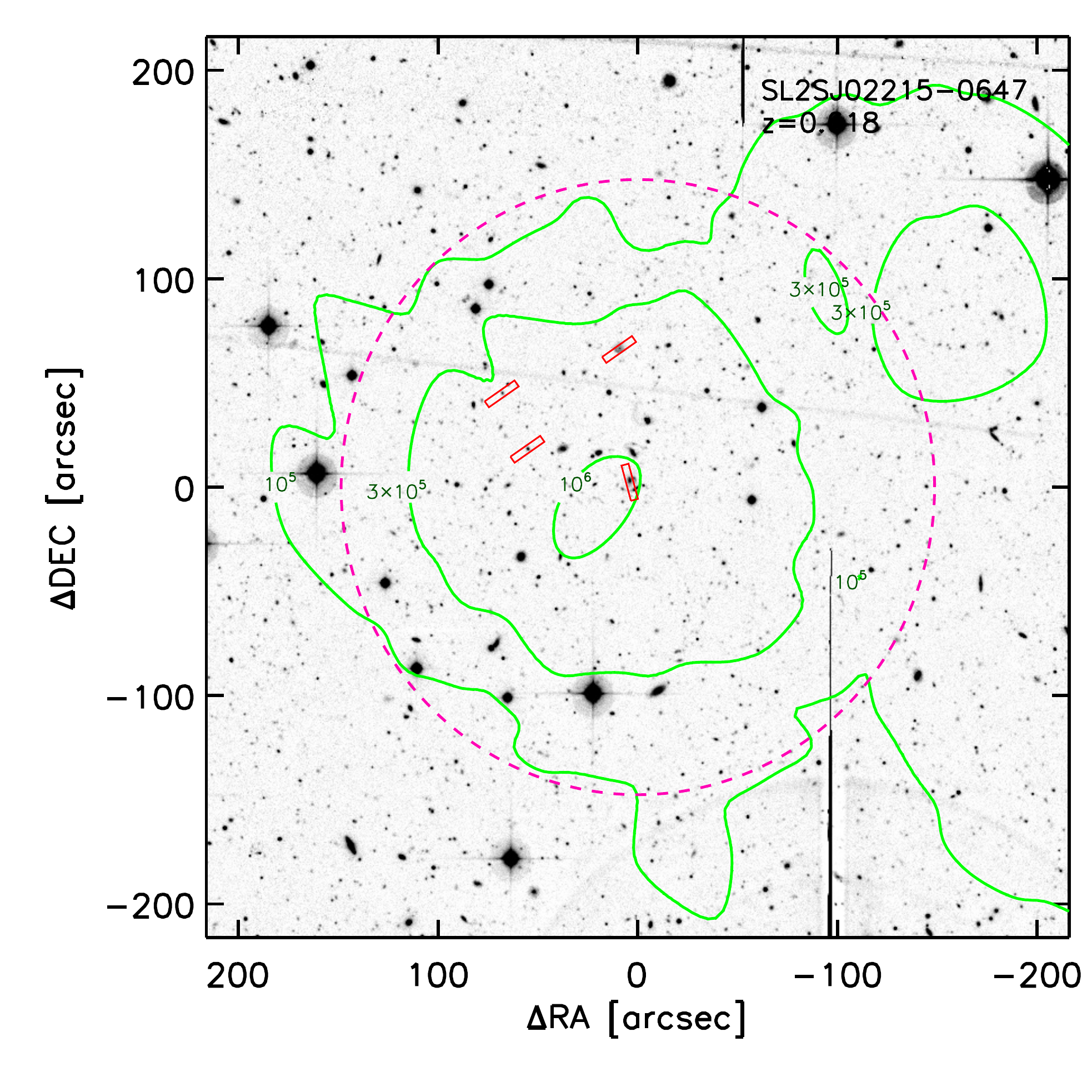}
\caption{Top panel: Optical spectra of the brightest confirmed member of SL2SJ02215-0647. The main absorption and emission lines used to determine the redshift of the main lens galaxy have been identified.
Bottom panel: Spatial distribution of galaxies in the group field of SL2SJ02215-0647. Red rectangles show the position of the spectroscopically confirmed members. The dashed magenta circle shows a circular aperture of radius $1\;Mpc$ at z=0.62. The contours are the same as in Fig. \ref{fig:spatial1}.}
\label{fig:spatial2}
\end{figure*}

\begin{figure*}[!htp]
\centering
\includegraphics[width=0.8\textwidth]{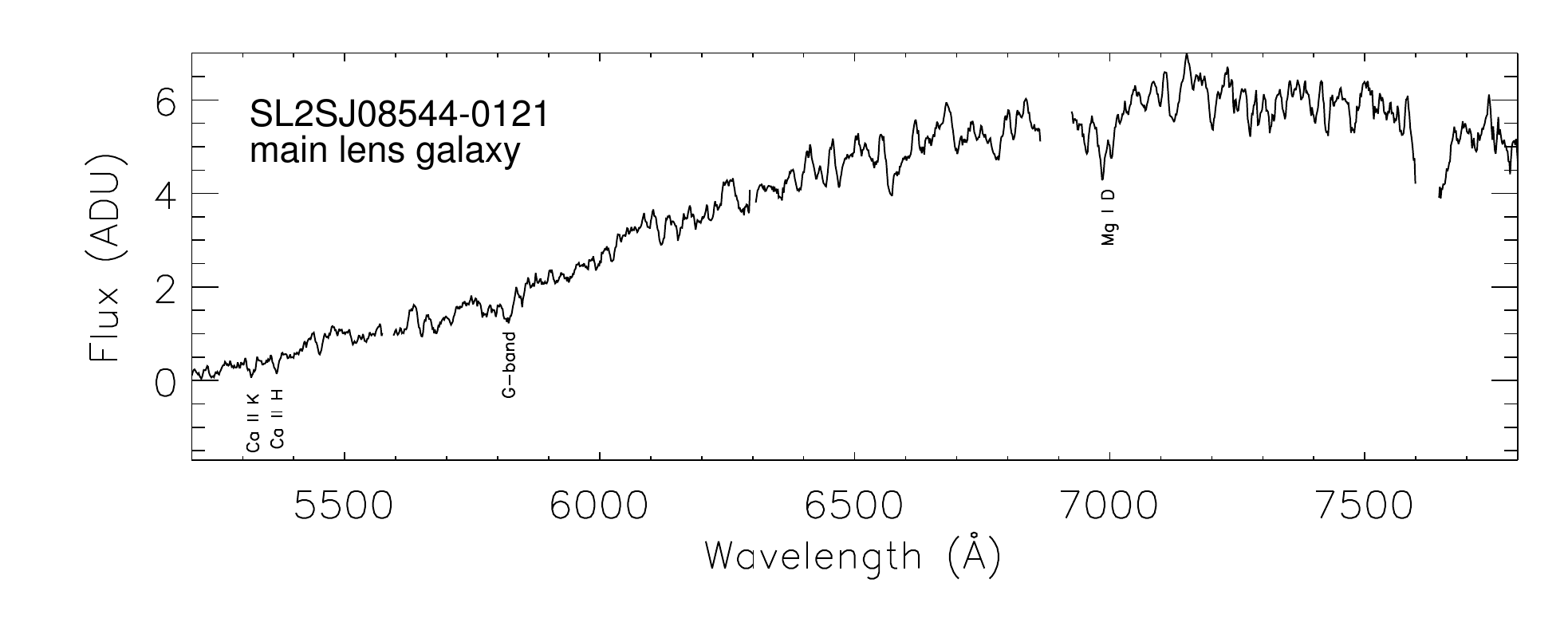}
\includegraphics[width=0.8\textwidth]{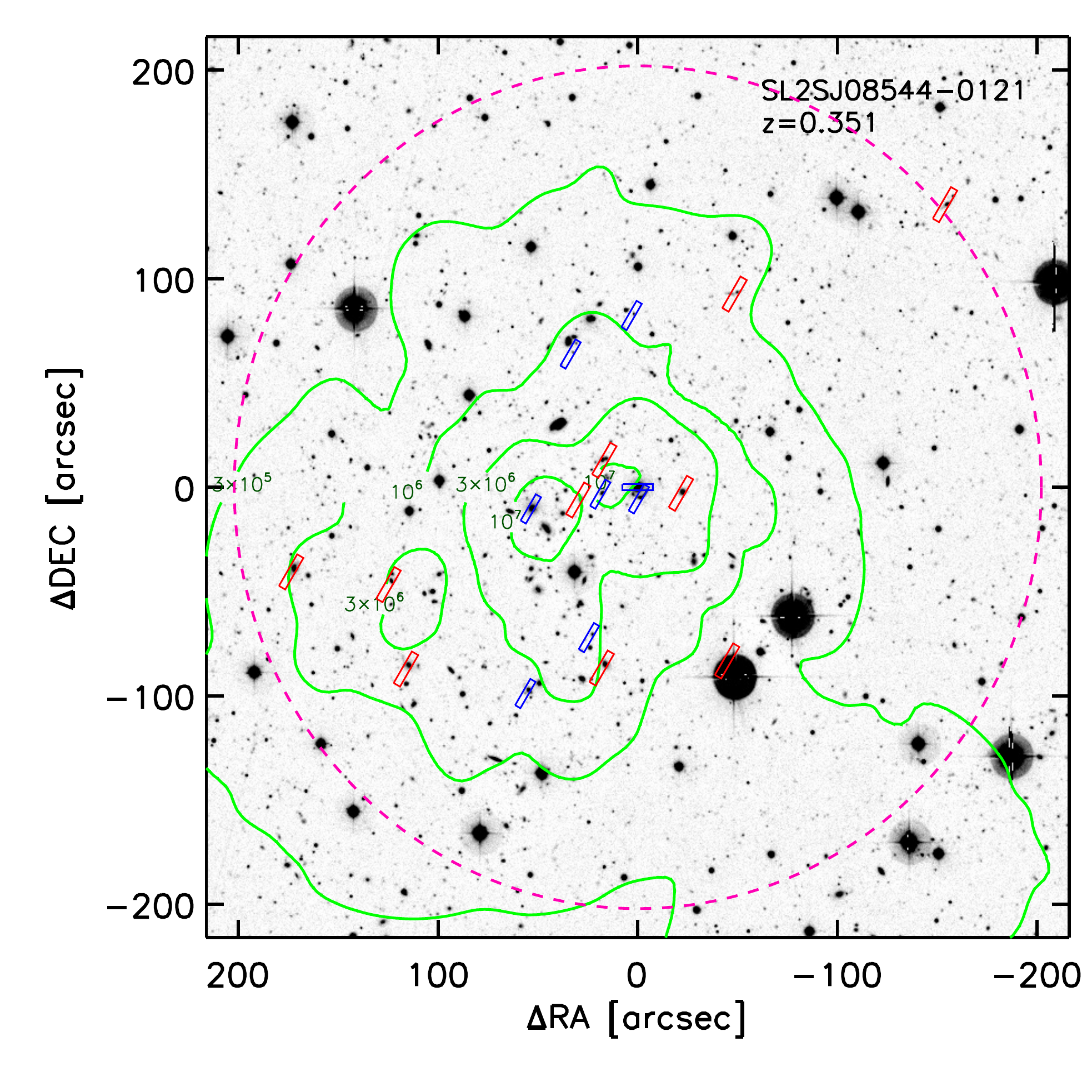}
\caption{Top panel: Optical spectra of the brightest confirmed member of SL2SJ08544-0121. The main absorption lines used to determine the redshift of the main lens galaxy have been identified.
Bottom panel: Spatial distribution of galaxies in the group field of SL2SJ08544-0121. Red rectangles show the position of the spectroscopically confirmed members. The dashed magenta circle shows a circular aperture of radius $1\;Mpc$ at z=0.35. The contours are the same as in Fig. \ref{fig:spatial1}.}
\label{fig:spatial3}
\end{figure*}

\begin{figure*}[!htp]
\centering
\includegraphics[width=0.8\textwidth]{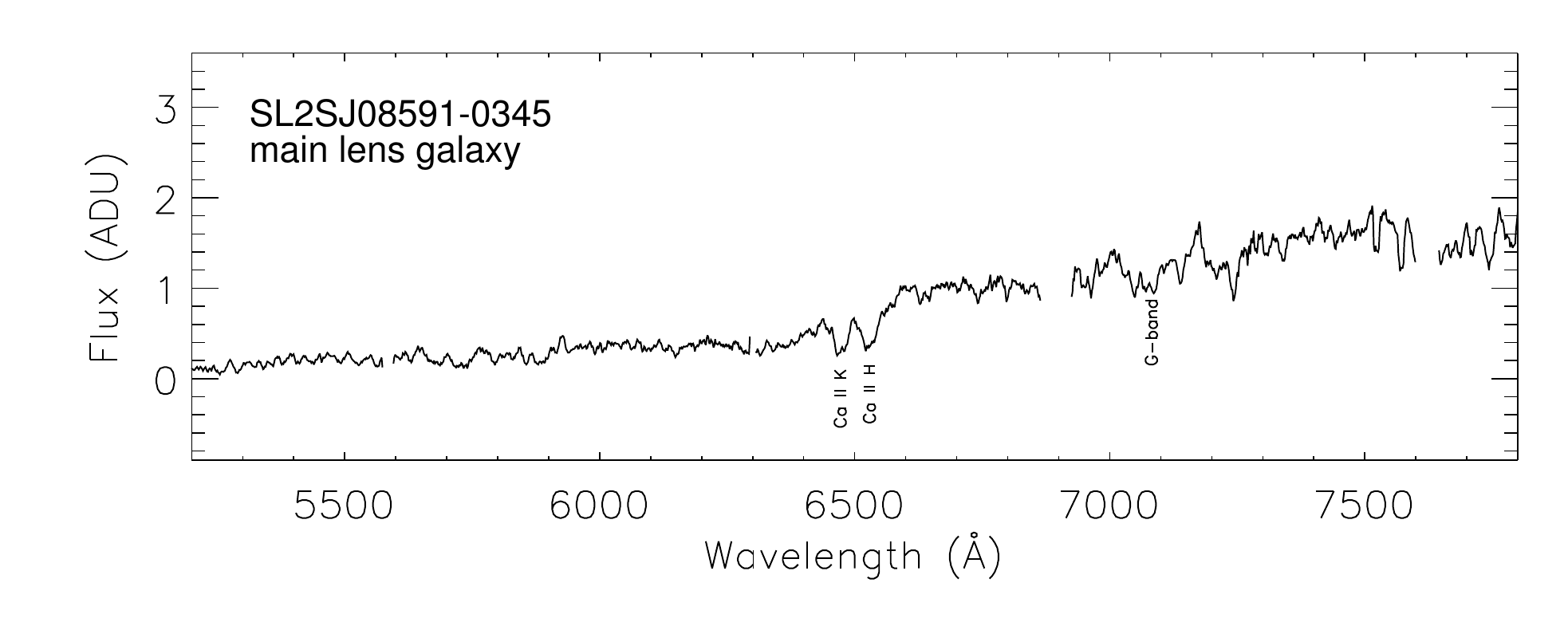}
\includegraphics[width=0.8\textwidth]{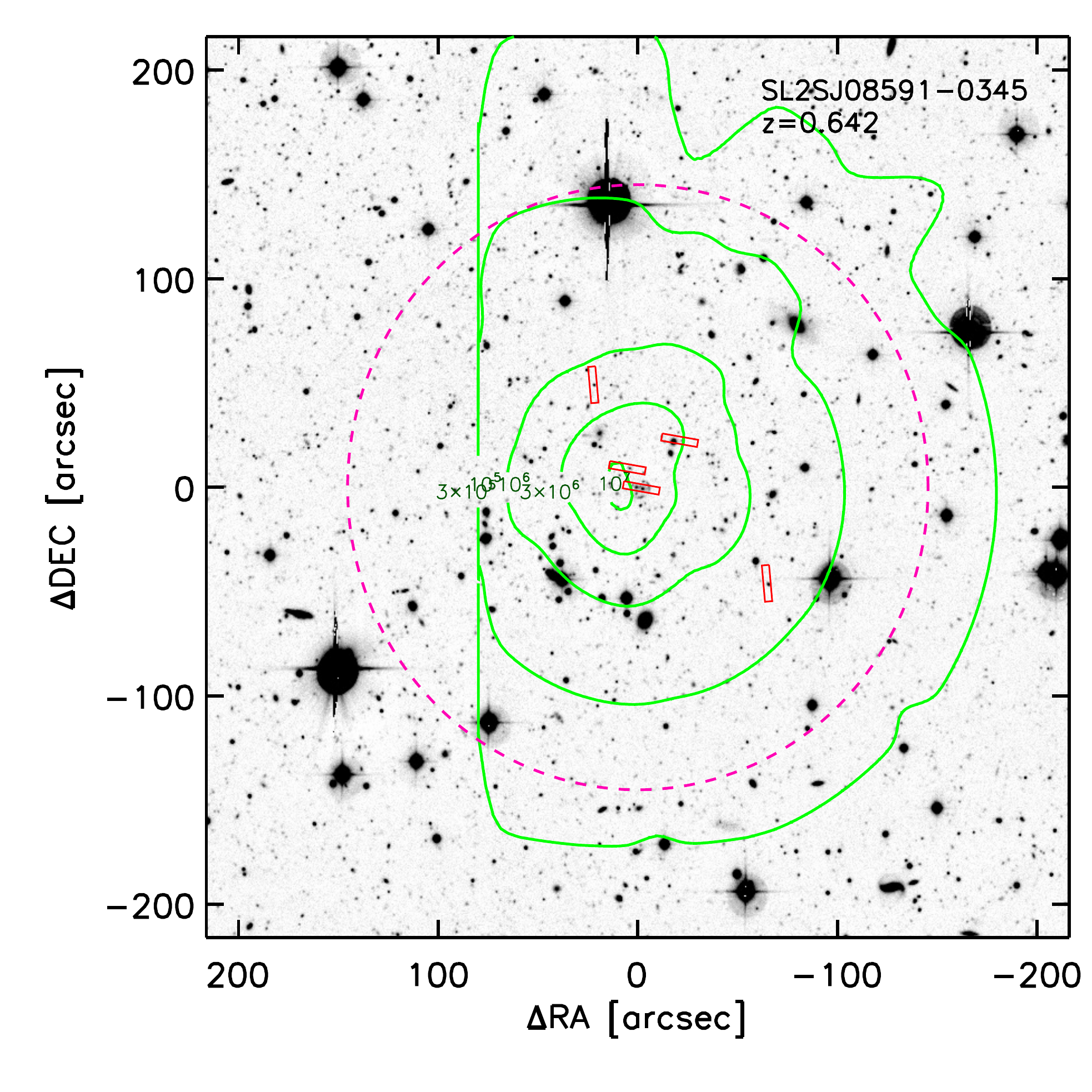}
\caption{Top panel: Optical spectra of the brightest confirmed member of SL2SJ08591-0345. The main absorption lines used to determine the redshift of the main lens galaxy have been identified.
Bottom panel: Spatial distribution of galaxies in the group field of SL2SJ08591-0345. Red rectangles show the position of the spectroscopically confirmed members. The dashed magenta circle shows a circular aperture of radius $1\;Mpc$ at z=0.64. The contours are the same as in Fig. \ref{fig:spatial1}.}
\label{fig:spatial2}
\end{figure*}

\begin{figure*}[!htp]
\centering
\includegraphics[width=0.8\textwidth]{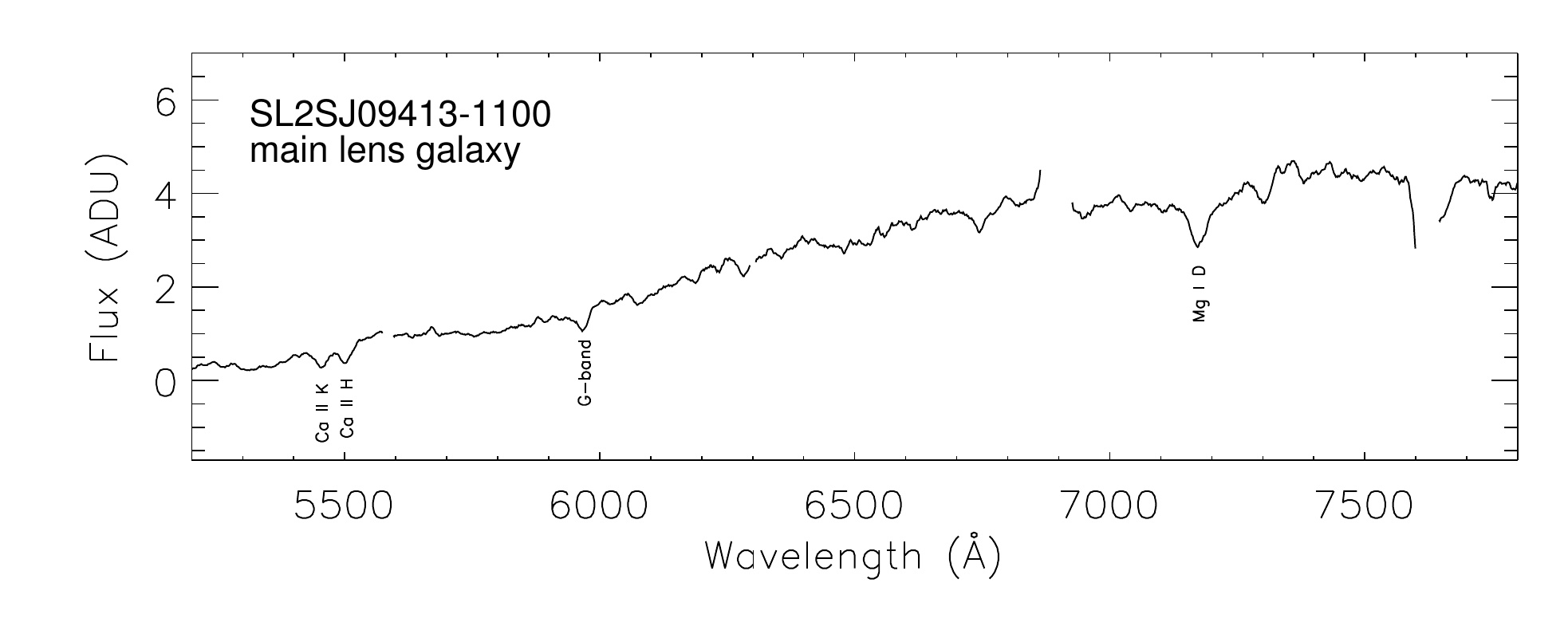}
\includegraphics[width=0.8\textwidth]{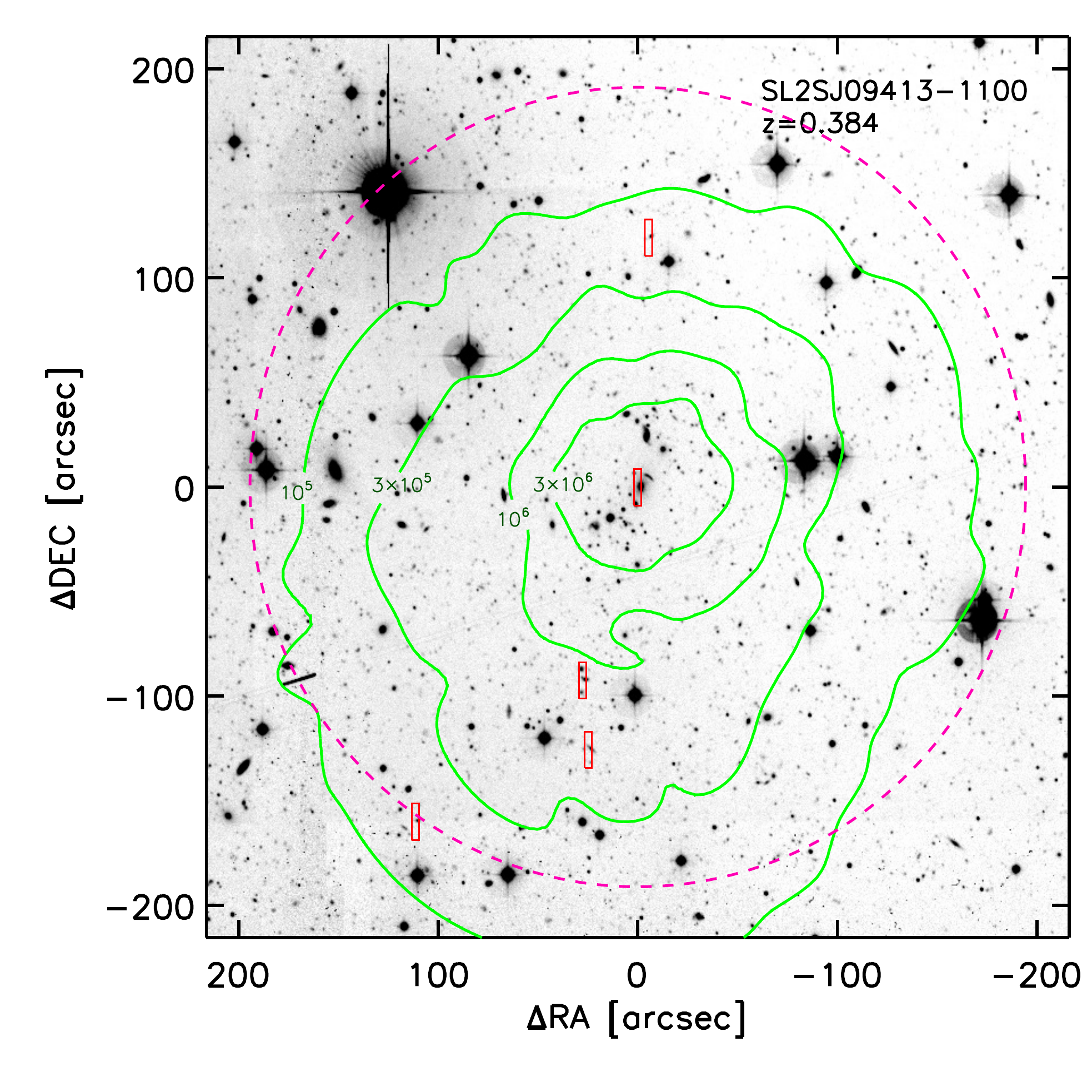}
\caption{Top panel: Optical spectra of the brightest confirmed member of SL2SJ09413-1100. The main absorption lines used to determine the redshift of the main lens galaxy have been identified.
Bottom panel: Spatial distribution of galaxies in the group field of SL2SJ09413-1100. Red rectangles show the position of the spectroscopically confirmed members. The dashed magenta circle shows a circular aperture of radius $1\;Mpc$ at z=0.39. The contours are the same as in Fig. \ref{fig:spatial1}.}
\label{fig:spatial2}
\end{figure*}

\end{appendix}
\clearpage

\begin{appendix}
\section{Summary of FORS2 masks and spectroscopic confirmed members for each group}

\longtab{1}{
\begin{longtable}{l l l r l l l }
\caption{\label{tab:mask}  Summary of group members.}\\
\hline\hline
\multicolumn{1}{c}{Group} & \multicolumn{1}{c}{Mask} & \multicolumn{1}{c}{Chip} & \multicolumn{1}{c}{Slit} & \multicolumn{1}{c}{RA} & \multicolumn{1}{c}{DEC} & \multicolumn{1}{c}{$z_{spec}$} \\
\hline
\endfirsthead
\caption{continued.}\\
\hline \hline
\multicolumn{1}{c}{Group} & \multicolumn{1}{c}{Mask} & \multicolumn{1}{c}{Chip} & \multicolumn{1}{c}{Slit} & \multicolumn{1}{c}{RA} & \multicolumn{1}{c}{DEC} & \multicolumn{1}{c}{$z_{spec}$} \\
\hline
\endhead
\hline
\endfoot
\multirow{16}{*}{SL2SJ02140-0535} & M012 & CHIP1 &  2 &  33.550777 &  -5.551144 &   0.444 \\
& M012 & CHIP1 &  5 &  33.536942 &  -5.582868 &   0.443 \\
& M012 & CHIP1 &  7 &  33.533779 &  -5.592632 &   0.445 \\
& M012 & CHIP1 &  9 &  33.530430 &  -5.594814 &   0.447 \\
& M012 & CHIP1 & 10 &  33.531372 &  -5.569443 &   0.449 \\
& M012 & CHIP1 & 15 &  33.519188 &  -5.601521 &   0.446 \\
& M012 & CHIP1 & 19 &  33.515137 &  -5.577593 &   0.444 \\
& M012 & CHIP1 & 21 &  33.512367 &  -5.573329 &   0.444 \\
& M012 & CHIP1 & 25 &  33.500538 &  -5.558484 &   0.446 \\
& M012 & CHIP1 & 29 &  33.484375 &  -5.623324 &   0.444 \\
& M012 & CHIP1 & 33 &  33.479259 &  -5.613928 &   0.444 \\
& M010 & CHIP1 &  1 &  33.546135 &  -5.607511 &   0.444 \\
& M010 & CHIP1 &  2 &  33.540424 &  -5.584474 &   0.443 \\
& M010 & CHIP1 &  9 &  33.512676 &  -5.596797 &   0.447 \\
& M010 & CHIP2 &  5 &  33.548912 &  -5.616460 &   0.443 \\
& LRIS & $-$ & $-$ &  33.533501 &  -5.591930 &   0.445 \\
\hline
\multirow{7}{*}{SL2SJ02141-0405} & M014\_1 & CHIP1 &  3 &  33.552521 &  -4.079994 &   0.605\\
& M014\_1 & CHIP1 &  4 &  33.551716 &  -4.081113 &   0.610 \\
& M014\_1 & CHIP1 &  6 &  33.548416 &  -4.083041 &   0.605 \\
& M014\_1 & CHIP1 &  7 &  33.547485 &  -4.083448 &   0.608 \\
& M014\_1 & CHIP1 & 22 &  33.522072 &  -4.091465 &   0.611 \\
& M014\_1 & CHIP1 & 27 &  33.512501 &  -4.099202 &   0.610 \\
& M014\_1 & CHIP2 & 10 &  33.578384 &  -4.060563 &   0.610 \\
\hline
\multirow{6}{*}{SL2SJ02180-0515A} & M014\_2 & CHIP1 &  9 &  34.536358 &  -5.252573 &   0.644 \\
& M014\_2 & CHIP1 & 13 &  34.561638 &  -5.247695 &   0.643 \\
& M014\_2 & CHIP1 & 14 &  34.564087 &  -5.246713 &   0.643 \\
& M014\_2 & CHIP2 & 10 &  34.538578 &  -5.285957 &   0.642 \\
& M014\_3 & CHIP1 &  3 &  34.542030 &  -5.259311 &   0.643 \\
& M014\_3 & CHIP1 &  6 &  34.531658 &  -5.259928 &   0.644 \\
\hline
\multirow{5}{*}{SL2SJ02180-0515B} & M014\_2 & CHIP1 &  5 &  34.531971 &  -5.261622 &   0.646 \\
& M014\_2 & CHIP1 &  8 &  34.546001 &  -5.258788 &   0.648 \\
& M014\_2 & CHIP2 &  9 &  34.518215 &  -5.283539 &   0.648 \\
& M014\_3 & CHIP1 &  5 &  34.532192 &  -5.260464 &   0.646 \\
& M014\_3 & CHIP1 &  7 &  34.530411 &  -5.260028 &   0.648 \\
\hline
\multirow{4}{*}{SL2SJ02215-0647} & M016\_1 & CHIP1 &  4 &  35.477470 &  -6.788154 &   0.617 \\
& M016\_1 & CHIP1 &  5 &  35.481045 &  -6.780771 &   0.617 \\
& M016\_1 & CHIP1 & 12 &  35.464710 &  -6.774853 &   0.619 \\
& M016\_2 & CHIP1 &  3 &  35.463230 &  -6.792549 &   0.617 \\
\pagebreak
\multirow{8}{*}{SL2SJ08544-0121A} & M005 & CHIP1 &  4 & 133.708832 &  -1.363143 &   0.353 \\
& M005 & CHIP1 &  7 & 133.699173 &  -1.361177 &   0.351 \\
& M005 & CHIP1 &  8 & 133.693802 &  -1.361807 &   0.352 \\
& M005 & CHIP1 & 16 & 133.703308 &  -1.342510 &   0.353 \\
& M005 & CHIP1 & 21 & 133.694809 &  -1.337374 &   0.353 \\
& M005 & CHIP2 & 12 & 133.709595 &  -1.387739 &   0.352 \\
& M005 & CHIP2 & 17 & 133.700745 &  -1.380287 &   0.353 \\
& LRIS & $-$ & $-$ & 133.693954 &  -1.360260 &   0.352 \\
\hline
\multirow{10}{*}{SL2SJ08544-0121B} & M005 & CHIP1 &  2 & 133.681519 &  -1.383398 &   0.357 \\
& M005 & CHIP1 &  6 & 133.702240 &  -1.361963 &   0.357 \\
& M005 & CHIP1 & 10 & 133.698608 &  -1.356661 &   0.359 \\
& M005 & CHIP1 & 11 & 133.687943 &  -1.361066 &   0.358 \\
& M005 & CHIP1 & 25 & 133.680481 &  -1.334525 &   0.360 \\
& M005 & CHIP1 & 33 & 133.651199 &  -1.322610 &   0.356 \\
& M005 & CHIP2 &  9 & 133.726212 &  -1.384477 &   0.356 \\
& M005 & CHIP2 & 11 & 133.742172 &  -1.371531 &   0.355 \\
& M005 & CHIP2 & 13 & 133.728638 &  -1.373227 &   0.354 \\
& M005 & CHIP2 & 16 & 133.698975 &  -1.384341 &   0.358 \\
\hline
\multirow{5}{*}{SL2SJ08591-0345} & M002\_1 & CHIP1 &  5 & 134.812363 &  -3.751204 &   0.644 \\
& M002\_1 & CHIP1 &  6 & 134.810410 &  -3.753887 &   0.645 \\
& M002\_1 & CHIP1 &  9 & 134.805069 &  -3.747538 &   0.642 \\
& M002\_2 & CHIP1 &  1 & 134.817093 &  -3.740146 &   0.644 \\
& M002\_2 & CHIP1 & 19 & 134.792923 &  -3.766603 &   0.639 \\
\hline
\multirow{5}{*}{SL2SJ09413-1100} & M001 & CHIP1 &  5 & 145.394897 & -11.015464 &   0.386 \\
& M001 & CHIP1 & 16 & 145.393402 & -10.982292 &   0.386 \\
& M001 & CHIP2 &  5 & 145.425903 & -11.059779 &   0.383 \\
& M001 & CHIP2 &  8 & 145.401794 & -11.050273 &   0.385 \\
& M001 & CHIP2 & 10 & 145.402603 & -11.041045 &   0.386 \\
\end{longtable}
}
\end{appendix}

\end{document}